\documentclass[twocolumn,showkeys,showpacs,preprintnumbers,prd,superscriptaddress,nofootinbib]{revtex4-1}
\usepackage{graphicx,epsf,bm,amsmath,amsfonts,amssymb,epstopdf,natbib,hyperref,color,verbatim,multirow,bm}
\hypersetup{colorlinks=true,urlcolor=blue,citecolor=blue,linkcolor=blue,menucolor=blue,anchorcolor=blue,filecolor=blue}

\date{\today}
            
\begin{document}

\title{Seven hints that early-time new physics alone\\is not sufficient to solve the Hubble tension}

\author{Sunny Vagnozzi}
\email{sunny.vagnozzi@unitn.it}
\affiliation{Department of Physics, University of Trento, Via Sommarive 14, 38123 Povo (TN), Italy}
\affiliation{Trento Institute for Fundamental Physics and Applications (TIFPA)-INFN, Via Sommarive 14, 38123 Povo (TN), Italy}

\begin{abstract}
\noindent The Hubble tension has now grown to a level of significance which can no longer be ignored and calls for a solution which, despite a huge number of attempts, has so far eluded us. Significant efforts in the literature have focused on early-time modifications of $\Lambda$CDM, introducing new physics operating prior to recombination and reducing the sound horizon. In this opinion paper I argue that early-time new physics \textit{alone} will always fall short of fully solving the Hubble tension. I base my arguments on seven independent hints, related to 1) the ages of the oldest astrophysical objects, 2) considerations on the sound horizon-Hubble constant degeneracy directions in cosmological data, 3) the important role of cosmic chronometers, 4) a number of ``descending trends'' observed in a wide variety of low-redshift datasets, 5) the early integrated Sachs-Wolfe effect as an early-time consistency test of $\Lambda$CDM, 6) early-Universe physics insensitive and uncalibrated cosmic standard constraints on the matter density, and finally 7) equality wavenumber-based constraints on the Hubble constant from galaxy power spectrum measurements. I argue that a promising way forward should ultimately involve a combination of early- and late-time (but non-local -- in a cosmological sense, i.e.\ at high redshift) new physics, as well as local (i.e.\ at $z \sim 0$) new physics, and I conclude by providing reflections with regards to potentially interesting models which may also help with the $S_8$ tension.
\end{abstract}

\maketitle

\section{Introduction}
\label{sec:introduction}

The concordance $\Lambda$CDM model has been extremely successful in describing cosmological and astrophysical observations across a wide range of times and scales, such as anisotropies in the Cosmic Microwave Background (CMB), the clustering of the large-scale structure (LSS), weak lensing of the CMB and the LSS (cosmic shear), the magnitude-redshift relation of distant Type Ia Supernovae (SNeIa), and light element abundances~\cite{SupernovaSearchTeam:1998fmf,SupernovaCosmologyProject:1998vns,DES:2017qwj,Planck:2018vyg,SPT:2019fqo,ACT:2020gnv,eBOSS:2020yzd,KiDS:2020suj,Mossa:2020gjc,Brout:2022vxf}. Nonetheless, we know that $\Lambda$CDM cannot be the end of the story: at best, it is a phenomenological placeholder for our ignorance about the fundamental nature of dark matter (DM), dark energy (DE), and the origin of primordial perturbations. Theory considerations aside, possible observational hints for new physics have recently emerged, in the form of tensions between independent inferences of cosmological parameters assuming $\Lambda$CDM.

Among these discrepancies, a special position is held by the ``Hubble tension'': the mismatch between several early- and late-time inferences of the Hubble constant $H_0$. One of the most precise early-time inferences is obtained from CMB temperature, polarization, and lensing measurements by the \textit{Planck} satellite, which assuming $\Lambda$CDM yield $H_0=(67.36 \pm 0.54)\,{\rm km}/{\rm s}/{\rm Mpc}$~\cite{Planck:2018vyg}, improving to $(67.62 \pm 0.47)\,{\rm km}/{\rm s}/{\rm Mpc}$ once combined with Baryon Acoustic Oscillation (BAO) and Hubble flow SNeIa data~\cite{Planck:2018vyg}. In contrast, one of the most precise local measurements (model-independent in a cosmological sense, although dependent on the underlying models for a variety of astrophysical effects), provided by the SH0ES team via a distance ladder making use of Cepheid-calibrated SNeIa, yields $(73.04 \pm 1.04)\,{\rm km}/{\rm s}/{\rm Mpc}$~\cite{Riess:2021jrx}. Depending on the dataset considered (see e.g.\ Refs.~\cite{FernandezArenas:2017dux,Riess:2019cxk,Wong:2019kwg,Freedman:2019jwv,Yuan:2019npk,Huang:2019yhh,Pesce:2020xfe,deJaeger:2020zpb,Schombert:2020pxm,Blakeslee:2021rqi} for other local measurements), the significance of the tension falls between $4\sigma$ and $6\sigma$, making it one of the most exciting open problems in cosmology (see Refs.~\cite{Verde:2019ivm,Riess:2019qba,DiValentino:2020zio,DiValentino:2021izs,Perivolaropoulos:2021jda,Schoneberg:2021qvd,Shah:2021onj,Abdalla:2022yfr,DiValentino:2022fjm,Hu:2023jqc} for reviews).

While systematics-based explanations (e.g.\ Refs.~\cite{Efstathiou:2020wxn,Mortsell:2021nzg,Mortsell:2021tcx,Freedman:2021ahq,Wojtak:2022bct}) are not yet completely excluded (but are admittedly becoming increasingly unlikely), the possibility of the Hubble tension calling for new physics, potentially related to the dark sector, is now taken very seriously, with a wide range of proposals having been put forward: with no claims as to completeness, see e.g.\ Refs.~\cite{Berezhiani:2015yta,Sola:2015wwa,Sola:2016jky,SolaPeracaula:2016qlq,DiValentino:2016hlg,Huang:2016fxc,Tram:2016rcw,Ko:2016uft,Karwal:2016vyq,Kumar:2016zpg,Xia:2016vnp,Ko:2016fcd,Chacko:2016kgg,Prilepina:2016rlq,Zhao:2017cud,Vagnozzi:2017ovm,Kumar:2017dnp,Benetti:2017gvm,Feng:2017nss,SolaPeracaula:2017esw,Zhao:2017urm,DiValentino:2017zyq,Dirian:2017pwp,DiValentino:2017iww,Sola:2017znb,Yang:2017zjs,Feng:2017mfs,Renk:2017rzu,Yang:2017alx,Buen-Abad:2017gxg,Yang:2017ccc,Raveri:2017jto,Sola:2016sbt,DiValentino:2017rcr,DiValentino:2017oaw,Khosravi:2017hfi,Santos:2017alg,Nunes:2017xon,Benetti:2017juy,Binder:2017lkj,Bolejko:2017fos,Feng:2017usu,Mortsell:2018mfj,Vagnozzi:2018jhn,Nunes:2018xbm,Poulin:2018zxs,Kumar:2018yhh,Yang:2018ubt,Bhattacharyya:2018fwb,Yang:2018euj,Poulin:2018dzj,Akarsu:2018aro,OColgain:2018czj,Yang:2018xlt,Banihashemi:2018oxo,Dutta:2018vmq,DEramo:2018vss,Guo:2018ans,Graef:2018fzu,Visinelli:2018utg,Yang:2018uae,El-Zant:2018bsc,Lin:2018nxe,Bengaly:2018xko,Yang:2018qmz,Bonilla:2018nau,Yang:2018xah,Banihashemi:2018has,SolaPeracaula:2018wwm,Poulin:2018cxd,Farrugia:2018mex,Carneiro:2018xwq,Kumar:2019gfl,Kenworthy:2019qwq,Kreisch:2019yzn,Pandey:2019plg,Martinelli:2019dau,Kumar:2019wfs,Vattis:2019efj,Pan:2019jqh,Akarsu:2019ygx,Li:2019san,Alexander:2019rsc,Agrawal:2019lmo,Yang:2019jwn,Hooper:2019gtx,Adhikari:2019fvb,Colgain:2019joh,Blinov:2019gcj,Keeley:2019esp,Lin:2019qug,Yang:2019qza,Agrawal:2019dlm,Li:2019yem,Martinelli:2019krf,Gelmini:2019deq,Rossi:2019lgt,DiValentino:2019exe,Yang:2019uzo,Archidiacono:2019wdp,Desmond:2019ygn,Yang:2019nhz,Nesseris:2019fwr,Pan:2019gop,Vagnozzi:2019ezj,Visinelli:2019qqu,Hardy:2019apu,Cai:2019bdh,Pan:2019hac,DiValentino:2019dzu,Xiao:2019ccl,Panpanich:2019fxq,DiValentino:2019ffd,Benetti:2019lxu,Dutta:2019pio,Ghosh:2019tab,SolaPeracaula:2019zsl,Escudero:2019gvw,Yan:2019gbw,Banerjee:2019kgu,Liu:2019dxr,Anchordoqui:2019yzc,Lunardini:2019zob,Yang:2019uog,DiValentino:2019jae,Niedermann:2019olb,Cheng:2019bkh,Berghaus:2019cls,Sakstein:2019fmf,Liu:2019awo,Anchordoqui:2019amx,Hart:2019dxi,Wang:2019isw,Alcaniz:2019kah,Gehrlein:2019iwl,Akarsu:2019hmw,Mazo:2019pzn,Ding:2019mmw,Ye:2020btb,Pan:2020zza,Amirhashchi:2020qep,Yang:2020zuk,Li:2020ybr,Beradze:2020pms,Perez:2020cwa,Lyu:2020lwm,Yang:2020uga,Choi:2020tqp,Dutta:2020jsy,Pan:2020bur,Sharov:2020bnk,Lucca:2020zjb,DAgostino:2020dhv,Hogg:2020rdp,Benevento:2020fev,Barker:2020gcp,Zumalacarregui:2020cjh,Hill:2020osr,Desmond:2020wep,Benisty:2020nql,Gomez-Valent:2020mqn,Akarsu:2020yqa,Ballesteros:2020sik,Blinov:2020uvz,Haridasu:2020xaa,Alestas:2020mvb,Jedamzik:2020krr,Braglia:2020iik,Chudaykin:2020acu,Izaurieta:2020xpk,Ballardini:2020iws,DiValentino:2020evt,Ivanov:2020mfr,Gogoi:2020qif,DiValentino:2020naf,Banerjee:2020xcn,Elizalde:2020mfs,Clark:2020miy,SolaPeracaula:2020vpg,Niedermann:2020dwg,Ivanov:2020ril,DAmico:2020ods,Elizalde:2020pps,DiValentino:2020kha,Gonzalez:2020fdy,Capozziello:2020nyq,Yang:2020ope,Sekiguchi:2020teg,Nunes:2020uex,Berechya:2020vcy,Benaoum:2020qsi,Calderon:2020hoc,Ye:2020oix,Flores:2020drq,Niedermann:2020qbw,Vazquez:2020ani,Akarsu:2020vii,Pogosian:2020ded,Smith:2020rxx,Murgia:2020ryi,Vagnozzi:2020rcz,Anchordoqui:2020djl,Hashim:2020sez,DiValentino:2020vnx,Odintsov:2020qzd,CarrilloGonzalez:2020oac,Das:2020xke,Braglia:2020auw,Adi:2020qqf,Banihashemi:2020wtb,RoyChoudhury:2020dmd,Mavromatos:2020kzj,Cai:2020tpy,Haridasu:2020pms,Shimon:2020mvl,Brinckmann:2020bcn,Alestas:2020zol,Yin:2020dwl,Moshafi:2020rkq,Yang:2021hxg,Seto:2021xua,Sinha:2021tnr,Gao:2021xnk,Cai:2021wgv,DiValentino:2021zxy,Marra:2021fvf,SolaPeracaula:2021gxi,Kumar:2021eev,Ren:2021tfi,Escudero:2021rfi,LeviSaid:2021yat,Nojiri:2021dze,Ye:2021nej,Elizalde:2021kmo,Dhawan:2021mel,Acquaviva:2021jov,Yang:2021oxc,Paul:2021ewd,Geng:2021jso,Hashim:2021pkq,Alestas:2021nmi,Fung:2021fcj,Mavromatos:2021urx,Teng:2021cvy,Thiele:2021okz,Zhou:2021xov,LinaresCedeno:2021aqk,Gu:2021lni,Lucca:2021dxo,Krishnan:2021dyb,Okamatsu:2021jil,Jin:2021pcv,Krishnan:2021jmh,Dinda:2021ffa,Adil:2021zxp,deAraujo:2021cnd,BeltranJimenez:2021wbq,Karwal:2021vpk,Lucca:2021eqy,Bag:2021cqm,Jiang:2021bab,Ghosh:2021axu,Nunes:2021zzi,Gomez-Valent:2021cbe,Li:2021htg,Hart:2021kad,Pogosian:2021mcs,Cyr-Racine:2021oal,Ye:2021iwa,Normann:2021bjy,Anchordoqui:2021gji,Staicova:2021ntm,Chang:2021yog,Drees:2021rsg,Rashkovetskyi:2021rwg,Qiu:2021cww,Liu:2021mkv,Rezaei:2021qwd,Nilsson:2021ute,Akarsu:2021fol,Luongo:2021nqh,Mawas:2021tdx,Perivolaropoulos:2021bds,Hill:2021yec,Feng:2021ipq,Poulin:2021bjr,Theodoropoulos:2021hkk,Hogg:2021yiz,Drepanou:2021jiv,Khosravi:2021csn,Aghababaei:2021gxe,Petronikolou:2021shp,Guo:2021rrz,DiValentino:2021rjj,Castello:2021uad,Bansal:2021dfh,Alestas:2021luu,DeSimone:2021rus,Clark:2021hlo,Aloni:2021eaq,Luu:2021yhl,Tiwari:2021ikr,Gariazzo:2021qtg,Benetti:2021uea,Belgacem:2021ieb,Duan:2021whx,Dialektopoulos:2021wde,Gomez-Valent:2021hda,Corona:2021qxl,Niedermann:2021ijp,Niedermann:2021vgd,DiValentino:2021imh,Renzi:2021xii,Takahashi:2021bti,Spallicci:2021kye,Akarsu:2021max,Saridakis:2021xqy,McDonough:2021pdg,Sen:2021wld,LaPosta:2021pgm,Greene:2021shv,Herold:2021ksg,Sakr:2021nja,Cao:2021zpf,Solomon:2022qqf,Colaco:2022aut,Banihashemi:2022vfv,Alestas:2022xxm,Wang:2022jpo,Odintsov:2022eqm,Perivolaropoulos:2022vql,Roy:2022fif,Dainotti:2022bzg,Heisenberg:2022lob,Hazra:2022rdl,Wang:2022rvf,Heisenberg:2022gqk,Qi:2022sxm,Sharma:2022ifr,Lee:2022cyh,Benisty:2022psx,Aboubrahim:2022gjb,Hoshiya:2022ady,Sabla:2022xzj,Heeck:2022znj,Benevento:2022cql,Smith:2022hwi,Ye:2022afu,Mazo:2022auo,Jin:2022qnj,Cai:2022dkh,Jiang:2022uyg,Mehrabi:2022mnr,Ren:2022aeo,Anchordoqui:2022gmw,Adhikari:2022moo,Simon:2022ftd,Nunes:2022bhn,Harko:2022unn,Davari:2022uwd,Qi:2022kfg,Fondi:2022tfp,Hu:2022kes,Gomez-Valent:2022hkb,Jin:2022tdf,Archidiacono:2022iuu,Berghaus:2022cwf,Yusofi:2022vsg,Carneiro:2022rmw,Ye:2022efx,Kumar:2022vee,Camarena:2022iae,Alvi:2022aam,Garcia-Arroyo:2022vvy,Aljaf:2022fbk,Schiavone:2022shz,Vagnozzi:2022moj,Nojiri:2022ski,Sandoval-Orozco:2022not,Kojima:2022fgo,Yao:2020hkw,MohseniSadjadi:2022pfz,Oikonomou:2022yle,Hernandez-Jimenez:2022daw,Schoneberg:2022grr,Seto:2022xgx,Reeves:2022aoi,Zhai:2022zif,Joseph:2022jsf,Aluri:2022hzs,Adil:2022hkj,Trivedi:2022bis,Akarsu:2022lhx,Wang:2022ssr,Rezaei:2022bkb,Gomez-Valent:2022bku,ElBourakadi:2022rgv,Moshafi:2022mva,Simon:2022adh,Buen-Abad:2022kgf,Odintsov:2022umu,Rezazadeh:2022lsf,Perivolaropoulos:2022khd,Escudero:2022rbq,Banerjee:2022ynv,Cruz:2022oqk,Oikonomou:2022tjm,Naidoo:2022rda,Wang:2022nap,Kumar:2022nvf,deSa:2022hsh,DiValentino:2022oon,Secco:2022kqg,Schoneberg:2022ggi,Jusufi:2022mir,Jiang:2022qlj,Zhou:2022wey,Zhao:2022bpd,Yang:2022kho,Cardona:2022pwm,Herold:2022iib,Holm:2022kkd,Akarsu:2022typ,Cai:2022dov,Pan:2022qrr,Gangopadhyay:2022bsh,Kuzmichev:2022kqb,Schiavone:2022wvq,Bernardo:2022ggl,Bansal:2022qbi,Lee:2022gzh,DiValentino:2022eot,Lin:2022phm,Haridasu:2022dyp,Brinckmann:2022ajr,Gao:2022ahg,deJesus:2022pux,Perivolaropoulos:2023iqj,Bouche:2023xjw,Brissenden:2023yko,Dahmani:2023goa,Khodadi:2023ezj,Bernui:2023byc,Thakur:2023ofa,Mandal:2023bzo,Jin:2023zhi,Kumar:2023bqj,Tiwari:2023jle,Dainotti:2023yrk,Bassi:2023ymf,Ben-Dayan:2023rgt,deCruzPerez:2023wzd,Ballardini:2023mzm,Giani:2023tai,Reboucas:2023rjm,Cruz:2023cxy,CarrilloGonzalez:2023lma,Goldstein:2023gnw,Yao:2023ybs,Borges:2023xwx,Das:2023npl,Staicova:2023jic,Jiang:2023bsz,Hogas:2023vim,Jusufi:2023ayv,Hill:2023wda,Nojiri:2023mvi,SolaPeracaula:2023swx,Sakr:2023mao,Sakr:2023hrl,Cruz:2023lmn,Perivolaropoulos:2023tdt,Allali:2023zbi,Dialektopoulos:2023jam,Dialektopoulos:2023dhb,Gomez-Valent:2023hov,Bisnovatyi-Kogan:2023frj,Odintsov:2023cli,Ye:2023zel,Jin:2023sfc,Takahashi:2023twt,Li:2023fdk,Buen-Abad:2023uva,Flores:2023nto,Ruchika:2023ugh,Greene:2023cro,Adil:2023exv,Montani:2023xpd,Schoneberg:2023rnx,Farhang:2023hen,Escamilla-Rivera:2023rop,Yadav:2023yyb,Lombriser:2023afv,Li:2023yaj,Singh:2023jxu,Nygaard:2023gel,Sarkar:2023vpn,Sengupta:2023yxh,Ferrari:2023qnh,Patil:2023rqy,Benaoum:2023ekz,Frion:2023xwq,Brax:2023tls,Liu:2023kce,DAgostino:2023cgx,Lu:2023uhc,Akarsu:2023mfb,Bousder:2023ida,Kable:2023bsg,Adil:2023ara,Das:2023rvg,Kalita:2023hqo,Peng:2023bik,Basilakos:2023kvk,Banerjee:2023teh,M:2023caa,Santana:2023gkx,Wei:2023avr,Hoerning:2023hks,Harada:2023afu} for examples of the rich variety of discussions in this context, as well as approaches adopted, with varying degrees of success. Cosmological solutions (involving new physics in the Hubble flow and not in the local Universe) usually feature modifications to $\Lambda$CDM either prior to recombination, or in the late Universe, which I shall refer to as \textit{early-time} and \textit{late-time} modifications respectively. It is now well understood that late-time models are less effective in addressing the Hubble tension, because of their worsened fit to BAO and Hubble flow SNeIa data when $H_0$ is increased, due to the fact that the sound horizon at baryon drag $r_d$ is not altered~\cite{Bernal:2016gxb,Addison:2017fdm,Lemos:2018smw,Aylor:2018drw,Schoneberg:2019wmt,Knox:2019rjx,Arendse:2019hev,Efstathiou:2021ocp,Cai:2021weh,Keeley:2022ojz}. In fact, the focus is now mostly towards early-time modifications, which aim to reduce $r_d$ by $\approx 7\%$ to accommodate a higher $H_0$ while keeping the angular size of the sound horizon to the CMB value, and not running afoul of late-time BAO and SNeIa constraints. Examples in this sense include but are not limited to models raising the pre-recombination expansion rate, or modifying the recombination history.

Nevertheless, it is fair to say that we remain far from a compelling solution to the Hubble tension. Leaving aside their theoretical motivation, none of the models proposed so far have succeeded in accommodating a higher $H_0$ while maintaining a good fit to all available data, or not worsening other tensions (e.g.\ the ``$S_8$ discrepancy''~\cite{DiValentino:2018gcu,DiValentino:2020vvd,Nunes:2021ipq,Huterer:2022dds,Sakr:2023bms}). Tongue-in-cheek, I would say that the statement ``the Hubble tension calls for early-time new physics'' may have been elevated to somewhat of a mantra in the community, and more often than not interpreted a bit too literally.~\footnote{Or with an added, uncalled for, adverb ``\textit{exclusively}'' between ``calls'' and ``for''.} My goal here is to argue that solving the Hubble tension will ultimately require more than just early-time new physics. I stress that this is not an original paper in a strict sense: to build my case, I will review results from a number of earlier works, at first glance perhaps unrelated to each other. When viewed more broadly, these results paint a coherent picture with a clear message: early-time new physics \textit{alone} is not sufficient to solve the Hubble tension. ``Alone'' is the key word here: there is no question that an important fraction of the ``tension-solving job'' needs to come from early-time physics (and that late-time new physics alone definitely cannot do the job), but the point is that something more, e.g.\ some amount of late-time or local new physics, is required. As a clarification, the terms ``local'' and ``non-local'' here are used not in the field theory sense of the principle of locality, but to distinguish physics in the local Universe ($z \sim 0$) from physics taking place at sufficiently high redshift. My case is built upon seven hints,~\footnote{It has been argued that the number of objects an average human can hold in short-term memory is $7 \pm 2$, a finding sometimes referred to as \textit{Miller's law}. Together with the many special and symbolic properties of the number seven, this is the reason behind my choice of focusing on seven hints.} some more theoretical in nature, others more data-driven, and others seemingly unrelated to the Hubble tension!

The rest of this paper is then organized as follows. The seven hints mentioned above are presented in Sec.~\ref{sec:sevenhints}. The latter is divided into seven subsections, each devoted to discussing one of these hints. In Sec.~\ref{sec:promising}, I speculate about promising model building directions towards solving the Hubble tension and possibly other tensions (such as the $S_8$ discrepancy) simultaneously, building upon the lessons learned from these seven hints. Finally, in Sec.~\ref{sec:conclusions} I draw concluding remarks.

\section{Seven hints}
\label{sec:sevenhints}

In what follows, I present the seven hints upon which my case is built. In hopes that this will help committing them to memory, I have developed an ``ABCDEFG'' mnemonic for the hints as follows:
\begin{itemize}
\item $\boldsymbol{\cal A}$ges of the oldest astrophysical objects;
\item $\boldsymbol{\cal B}$aryon Acoustic Oscillations sound horizon-Hubble constant degeneracy slope (compared to the same slope in the CMB);
\item $\boldsymbol{\cal C}$osmic chronometers;
\item $\boldsymbol{\cal D}$escending trends observed in a wide range of low-redshift data;
\item $\boldsymbol{\cal E}$arly integrated Sachs-Wolfe effect and the restrictions it imposes on early-time new physics;
\item $\boldsymbol{\cal F}$ractional matter density constraints from early-Universe physics insensitive and uncalibrated cosmic standards;
\item $\boldsymbol{\cal G}$alaxy power spectrum sound horizon- and equality wavenumber-based determinations of the Hubble constant.
\end{itemize}
In the following subsections, the hints will be presented in the above order, though I note that this is not necessarily the order which I would adopt to present them, say, in a talk.

\subsection{Ages of the oldest astrophysical objects}
\label{subsec:oao}

This first hint is, in principle, unrelated to the Hubble tension. The existence of old astrophysical objects (OAO) at high redshift has historically played an important role in cosmology, particularly with regards to hinting towards the existence of a cosmological constant-like component, and thus cosmic acceleration, way before the latter was actually discovered through SNeIa in 1998. The usefulness of OAO as a cosmological test is based on the following simple, incontrovertible fact: at any given redshift, the Universe must be \textit{at least as old} as the oldest objects it contains at that redshift. Being the age-redshift of the Universe a model-dependent function, this statement can be turned around into a cosmological test, which can be used to exclude those models (or parameters) leading to the Universe being paradoxically younger than its oldest objects.~\footnote{Prior to 1998, and particularly in the 1980s, the leading cosmological model envisaged an Einstein-de Sitter (EdS) Universe, with a vanishing cosmological constant and no spatial curvature. The existence of old galaxies at high redshift (in conjunction with increasingly precise measurements of $H_0$), seemingly older than the EdS Universe, gave rise to an important ``age crisis''~\cite{Jaffe:1995qu,Ostriker:1995su,Dunlop:1996mp}, whose resolution eventually came with the discovery of cosmic acceleration in 1998. In fact, compared to an EdS Universe with the same total energy density, a $\Lambda$CDM Universe where part of the matter content is replaced by dark energy naturally leads to an older Universe at any redshift, thus accommodating the otherwise puzzling OAO.}

\begin{figure*}
\begin{center}
\includegraphics[height=7.1cm]{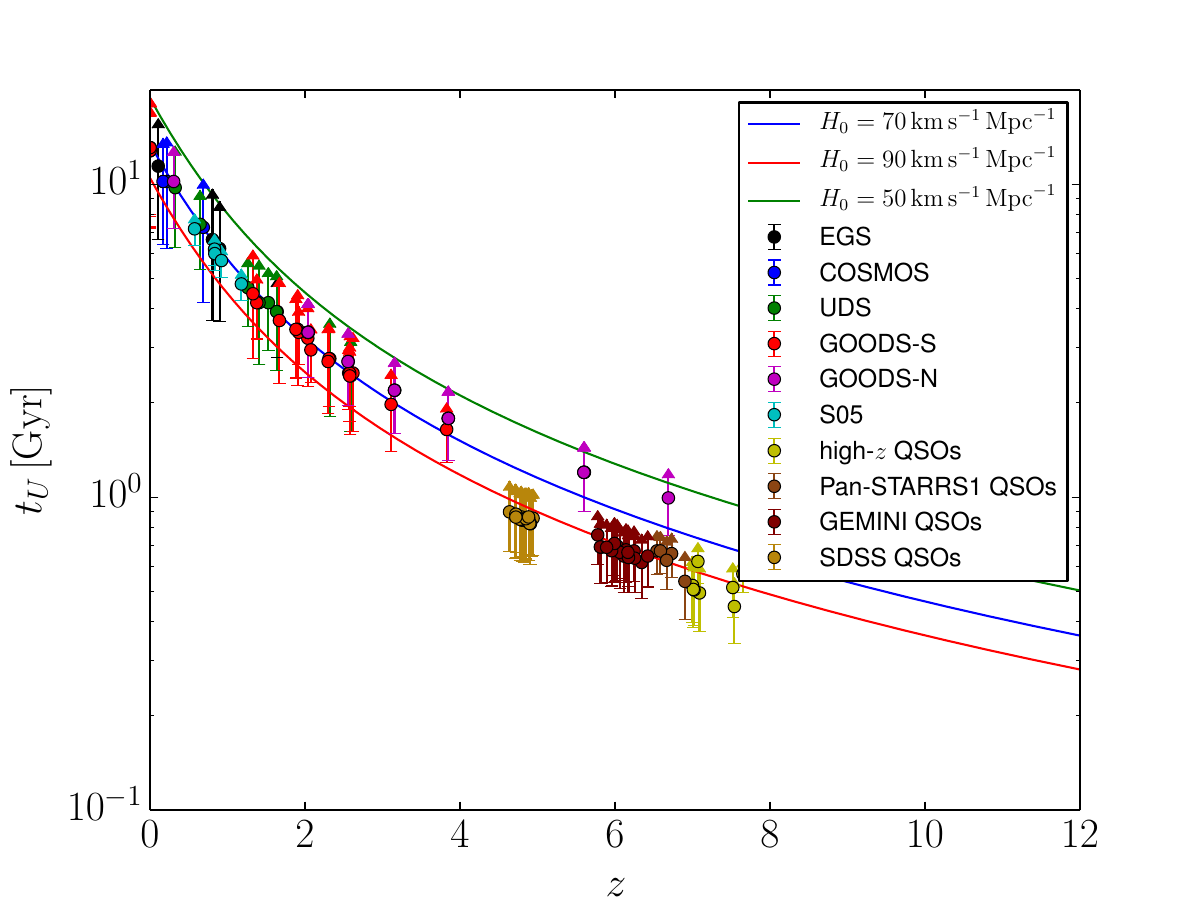}\,\includegraphics[height=6.5cm]{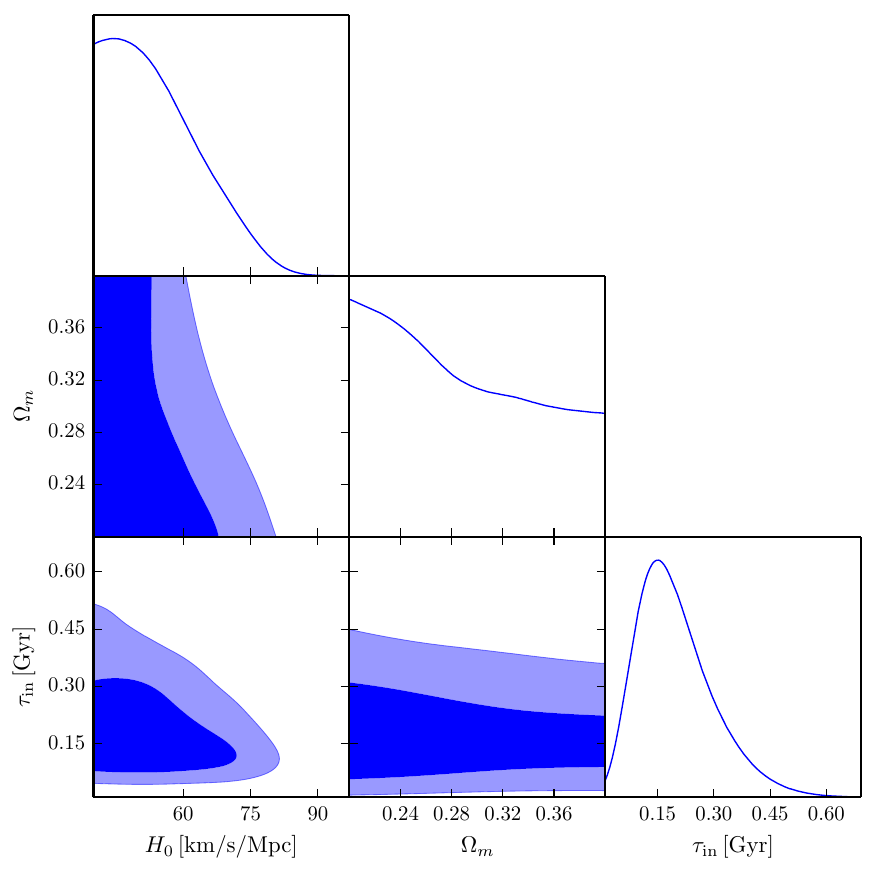}
\caption{\textit{Left panel}: age-redshift diagram of the OAO considered in Ref.~\cite{Vagnozzi:2021tjv}, with the curves showing the age-redshift relationship for three $\Lambda$CDM cosmologies with $\Omega_m=0.3$ and different values of $H_0$ as per the color coding. \textit{Right panel}: corner plot for $H_0$, $\Omega_m$, and $\tau_{\rm in}$ in light of the OAO age-redshift diagram. Reproduced from Fig.~1 and Fig.~2 of Vagnozzi, Pacucci \& Loeb, ``\textit{Implications for the Hubble tension from the ages of the oldest astrophysical objects}'', \textit{Journal of High Energy Astrophysics}, Volume 36, Pages 27-35, \href{https://www.sciencedirect.com/science/article/abs/pii/S2214404822000398}{doi:10.1016/j.jheap.2022.07.004}, published 4 August 2022~\cite{Vagnozzi:2021tjv}. \textcopyright\,(2022) Elsevier BV. Reproduced by permission of Elsevier and the first author. All rights reserved.}
\label{fig:oao}
\end{center}
\end{figure*}

To set the stage, the age of the Universe at any redshift, $t_{\cal U}(z)$, is given by the following integral:
\begin{eqnarray}
t_{\cal U}(z) &=& \int_{z}^{\infty} \frac{d\tilde{z}}{(1+\tilde{z})H(\tilde{z})} \nonumber \\
& \approx & \frac{977.8}{H_0\,[{\rm km}/{\rm s}/{\rm Mpc}]} \int_{z}^{\infty} \frac{d\tilde{z}}{(1+\tilde{z})E(\tilde{z})}\,{\rm Gyr}\,,
\label{eq:ageredshiftrelationship}
\end{eqnarray}
where $E(z) \equiv H(z)/H_0$ is the normalized expansion rate, and in what follows I will denote the age of the Universe today by $t_{\cal U} \equiv t_{\cal U}(z=0)$. Three comments are in order about Eq.~(\ref{eq:ageredshiftrelationship}):
\begin{enumerate}
\item the age of the Universe at any given redshift is self-evidently inversely proportional to the Hubble constant, $t_{\cal U}(z) \propto 1/H_0$;
\item the age integral picks up most of its contributions at late times, $z \lesssim 10$, since $E(z)$ increases faster than $(1+z)^{3/2}$ in the early Universe (see e.g.\ Ref.~\cite{Krishnan:2021dyb}) -- for all intents and purposes, it is therefore basically insensitive to pre-recombination (new) physics;~\footnote{Note that, although Eq.~(\ref{eq:ageredshiftrelationship}) is formally integrated up to $z \to \infty$, since in practice OAO can only form after recombination, the upper limit of the integral can be set to $z_{\star}$, the redshift of recombination. This prevents it from being sensitive to huge pre-BBN modifications to $E(z)$ which could make $t_U(z)$ arbitrarily large, although such modifications are not of interest in the discussion on the Hubble tension.}
\item it is possible for different cosmological models to lead to the same value of $t_{\cal U}$ (at $z=0$), while predicting a completely different evolution $t_{\cal U}(z)$ at high redshift.
\end{enumerate}
As myself, Pacucci and Loeb recently noted~\cite{Vagnozzi:2021tjv}, the first two points above indicate that OAO can play an important role in further uncovering the origin of the Hubble tension, if due to new physics. In particular, OAO can be used as an \textit{early-time-independent consistency test of late-time physics}, as follows~\cite{Vagnozzi:2021tjv}:
\begin{itemize}
\item create an OAO age-redshift catalog;
\item choose a given model for the late-time expansion (the ``null hypothesis'' which will be the subject of the consistency test);
\item impose (in a statistical sense) that the age of the Universe at any redshift within the chosen model exceeds the OAO ages -- given the previous point 1), this will lead to an upper limit on $H_0$;
\item the derived upper limit being in tension with local $H_0$ measurements would indicate an inconsistency in the chosen cosmological model and thus the need for at least some amount of late-time new physics (``new'' relative to the chosen model) -- conversely, absence of tension is at best an indication that there is no inconsistency yet, since it is in principle possible that OAO older than those present in the catalog may not yet have been identified.
\end{itemize}
Before moving on, I note that an important caveat to results involving OAO concerns the reliability of their ages, which are notoriously difficult to estimate. Nonetheless, significant progress is being made with regard to these issues (see e.g.\ Refs.~\cite{Pacucci:2019baa,Valcin:2021jcg,Borghi:2021zsr,Borghi:2021rft,Pacucci:2021ubg}), and more progress can be expected along these lines thanks to the recent launch of the James Webb Space Telescope. With these caveats in mind, in Ref.~\cite{Vagnozzi:2021tjv}, I led an analysis following the above recipe, selecting $\Lambda$CDM as the baseline model for the late-time expansion. In Ref.~\cite{Vagnozzi:2021tjv}, myself, Pacucci and Loeb constructed an age-redshift catalog of 114 OAO up to $z \approx 8$, considering both galaxies, most of which identified within the CANDELS observing program, and quasars (QSOs) identified within various surveys. Most galaxy ages were estimated via photometric spectral energy distribution fits, whereas the QSOs ages were estimated via Monte Carlo realizations of the growth model proposed by Pacucci \textit{et al.}~\cite{Pacucci:2017mcu}. I refer the reader to Section~3 of Ref.~\cite{Vagnozzi:2021tjv} for further details on the adopted galaxies and QSOs data, as well as on the age estimate methodology. The resulting OAO age-redshift diagram is shown in the left panel of Fig.~\ref{fig:oao}, alongside the predicted age-redshift relationship for three $\Lambda$CDM cosmologies with $\Omega_m=0.3$ and three different values of $H_0$.

To obtain (Bayesian) upper limit on $H_0$ from the OAO age-redshift diagram, Markov Chain Monte Carlo (MCMC) methodswere used, considering a 3-dimensional parameter space described by $\Omega_m$, $H_0$, and $\tau_{\rm in}$~\cite{Vagnozzi:2021tjv}. The latter is referred to as ``incubation time'' and accounts for the time elapsed between the Big Bang and the formation of the OAO, or in other words the fact that no OAO formed right at the Big Bang. From general considerations, we expect $\tau_{\rm in} \sim {\cal O}(0.1)\,{\rm Gyr}$~\cite{Jimenez:2019onw,Vagnozzi:2020dfn}, but to be as conservative as possible $\tau_{\rm in}$ was marginalized upon, imposing the so-called J19 prior, using the fitting function provided in Appendix~G of Ref.~\cite{Valcin:2020vav}: this approach was argued to be conservative, as it assumes that the OAO descend from the oldest generation of galaxies (had they been assumed to descended from a later generation of galaxies, more stringent but less conservative upper limits on $H_0$ would have been obtained), see Refs.~\cite{Jimenez:2019onw,Vagnozzi:2021tjv} for further details.

For the baseline analysis adopting flat priors on $H_0$ and $\Omega_m$, and the J19 prior on $\tau_{\rm in}$, regarded as the most balanced one in terms of equilibrium between conservative and aggressive assumptions, in Ref.~\cite{Vagnozzi:2021tjv} myself, Pacucci and Loeb found the 95\% confidence level (C.L.) upper limit $H_0<73.2\,{\rm km}/{\rm s}/{\rm Mpc}$: see the right panel of Fig.~\ref{fig:oao} for the corresponding contour plot. The concordance/discordance between this upper limit and the then-current local Cepheid-calibrated SNeIa distance ladder measurement was estimated to be at the $2.3\sigma$ level, certainly not high enough to be alarming, but still worthy of attention. This limit was found to be relatively stable against different analysis assumptions (particularly concerning priors on $\Omega_m$ and $\tau_{\rm in}$), with most of these assumptions actually resulting in more stringent upper limits on $H_0$, and hence stronger conclusions. Other analysis assumptions were further explored by Wei and Melia~\cite{Wei:2022plg}, and their effect on the resulting upper limits on $H_0$ were found to be small (see however 
Ref.~\cite{Costa:2023cmu}).

If we take this discrepancy seriously, what consequences follow? Assuming that the OAO ages have been correctly estimated, essentially only two further assumptions can be questioned: the validity of the $\Lambda$CDM model for the late-time expansion, and the validity of the local Cepheid-calibrated SNeIa estimate of $H_0$. Going down the first route requires introducing new (cosmological) physics which makes the Universe older at all redshifts: this can be achieved if new physics results in the expansion rate lowering relative to $\Lambda$CDM at $z>0$ (as in the case of a phantom-like component), in such a way to accommodate a higher $H_0$, of course to an extent which is compatible with BAO and Hubble flow SNeIa constraints -- more on this will be discussed in Sec.~\ref{sec:promising}. Going down the second route opens the increasingly unlikely door of systematics (already amply discussed in the literature, as noted in Sec.~\ref{sec:introduction}), or the possibility of new local physics affecting the distance ladder measurements, more precisely lowering it. While I will return to these points in much more detail in Sec.~\ref{sec:promising}, in the meantime I simply note that the possibilities of new late-time (cosmological) and local physics are not mutually exclusive, as the two can well act simultaneously (in addition to early-time new physics), the former to raise the cosmological $H_0$ estimate, the latter to lower the local $H_0$ estimate, bringing the two into better agreement. Of course, these results come with the caveats discussed earlier, pertaining to the difficulty of reliably estimating OAO ages. In particular, if mismodelled astrophysical or galaxy evolution effects cause the OAO ages to be systematically overestimated, this would directly result in a systematic underestimation of the upper limit on $H_0$, therefore artificially worsening the tension with local measurements.

Before moving on, a comment on related works is in order, as Ref.~\cite{Vagnozzi:2021tjv} was not to the only one to recently appreciate the role of galaxy ages in the quest towards arbitrating the Hubble tension. Similar arguments, albeit limited to the age of the Universe today $t_U$, were recently put forward~\cite{Jimenez:2019onw,Bernal:2021yli,Krishnan:2021dyb}. In essence these works pointed out that, should (lower limits on) the age of the Universe as obtained from old stars and galaxies confirm the high value of $t_U$ indicated by \textit{Planck} assuming $\Lambda$CDM (given the preferred low value of $H_0$, and the fact that $t_U \propto 1/H_0$), this would at the very least require introducing either some late-time new physics, or some local new physics: to put it differently, early-time new physics alone would not be sufficient to solve the Hubble tension in this case. A stronger version of this argument was put forward by Bernal \textit{et al.}~\cite{Bernal:2021yli}, who pointed out the key role of $t_U$ (alongside $\Omega_m$) in arbitrating the Hubble tension, highlighting the usefulness of ``cosmic triangles'' related to $H_0$, $t_U$, $\Omega_m$, and the sound horizon $r_d$, i.e.\ ternary plots simultaneously visualizing independent constraints on these parameters (which are over-constrained given the precision of current cosmological measurements). Similarly, Krishnan \textit{et al.}~\cite{Krishnan:2021dyb} instead argued that recent constraints on $t_U$ from old globular clusters, when analyzed in conjunction with a minimal parametrization for the late-time expansion while treating $r_d$ as a free parameter, indicate that early-time new physics can at best bring $H_0$ up to $\approx 71\,{\rm km}/{\rm s}/{\rm Mpc}$, confirming empirical findings that no early-time model so far has been able to do better than this. The results of Ref.~\cite{Vagnozzi:2021tjv}, which this first hint build upon, essentially constitute a stronger version of these arguments, built upon $t_U(z)$ rather than $t_U$ alone (recall that different cosmological models can lead to the same $t_U$ while predicting a completely different $t_U(z)$ evolution), and providing the first direct indication that OAO ages may indeed be in slight tension with local $H_0$ measurements. Finally, I note that the importance of OAO ages in arbitrating the Hubble tension was also discussed later by Borghi \textit{et al.}~\cite{Borghi:2021rft}, Wei and Melia~\cite{Wei:2022plg}, and Moresco \textit{et al.}~\cite{Moresco:2022phi}. For further related works, see Refs.~\cite{Boylan-Kolchin:2021fvy,Cimatti:2023gil,Jimenez:2023flo}, and Ref.~\cite{Capozziello:2023ewq} for a discussion in the context of the related look-back time quantity.

In closing, I think it is important to stress once more that this first hint is in principle completely unrelated to the Hubble tension. In essence, it constitutes of a consistency test whose failure indicates the need for some new physics at late times or on local scales, independently any assumed model for the early Universe. Of course, one would hope that this late-time and/or local new physics would go in the direction of helping with (or at the very least not worsen!) the Hubble tension, beyond the (lion's share?) of the job inevitably done by early-time new physics.

\subsection{BAO sound horizon-Hubble constant degeneracy slope}
\label{subsec:baodegeneracyslope}

The physics of acoustic oscillations, whose associated distance scale(s) provides a standard ruler through which distances at high redshift can be inferred, is cleanly imprinted in two classes of observations: in the CMB through the location of the acoustic peaks (particularly the first ones), and in the Baryon Acoustic Oscillation BAO peaks observed in correlators of tracers of the large-scale structure, such as galaxies. In the case of the CMB, the location of the acoustic peaks and their spacing determines $\theta_{\star}$, the angular size of the comoving sound horizon at recombination:
\begin{eqnarray}
\theta_{\star} \equiv \frac{r_{\star}}{D(z_{\star})}\,,
\label{eq:thetastarcmb}
\end{eqnarray}
where $r_{\star}$ is the comoving sound horizon at the epoch of recombination, occurring at redshift $z_{\star}$, and $D(z_{\star})$ is the comoving distance to recombination. Since $D(z_{\star})$ essentially only depends on the Hubble constant $H_0$ or equivalently the reduced Hubble constant $h \equiv H_0/(100\,{\rm km}/{\rm s}/{\rm Mpc})$, as well as the physical matter density parameter $\omega_m \equiv \Omega_mh^2$, measurements of $ \theta_{\star}$ can be used to infer $H_0$ once $\omega_m$ and $r_{\star}$ are known (or, in the case of $r_{\star}$, calibrated given a model).

\begin{figure*}
\begin{center}
\includegraphics[height=5.5cm]{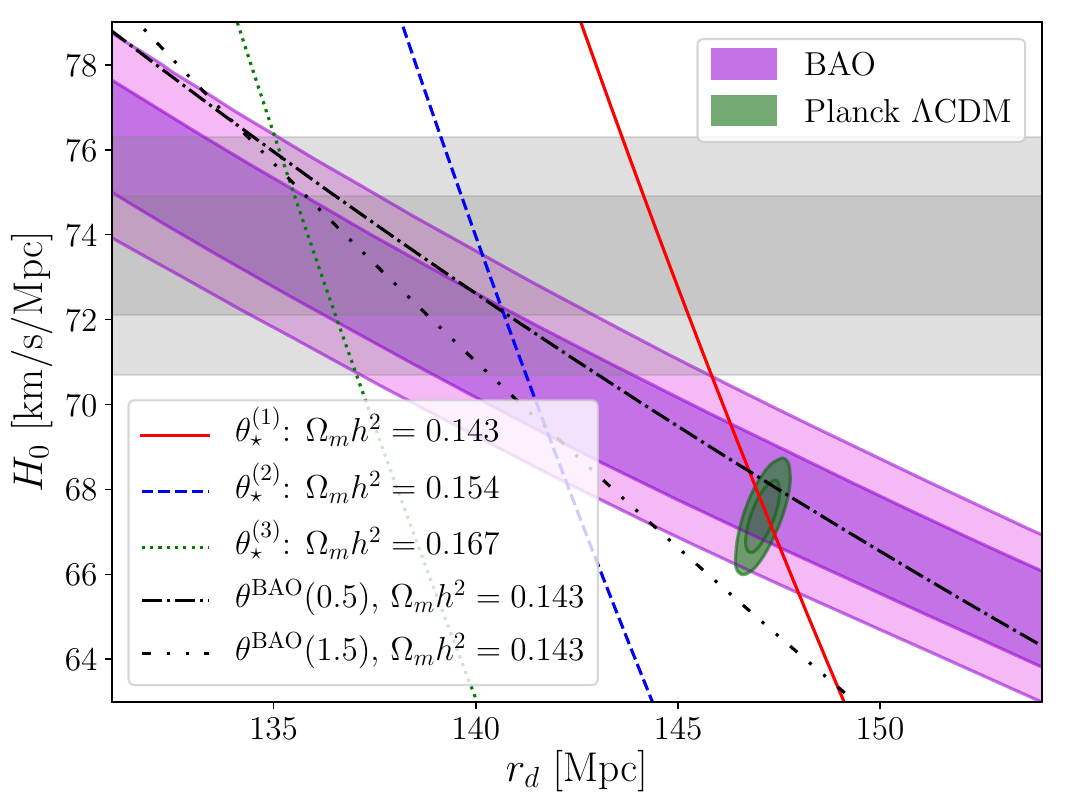}\,\includegraphics[height=5.5cm]{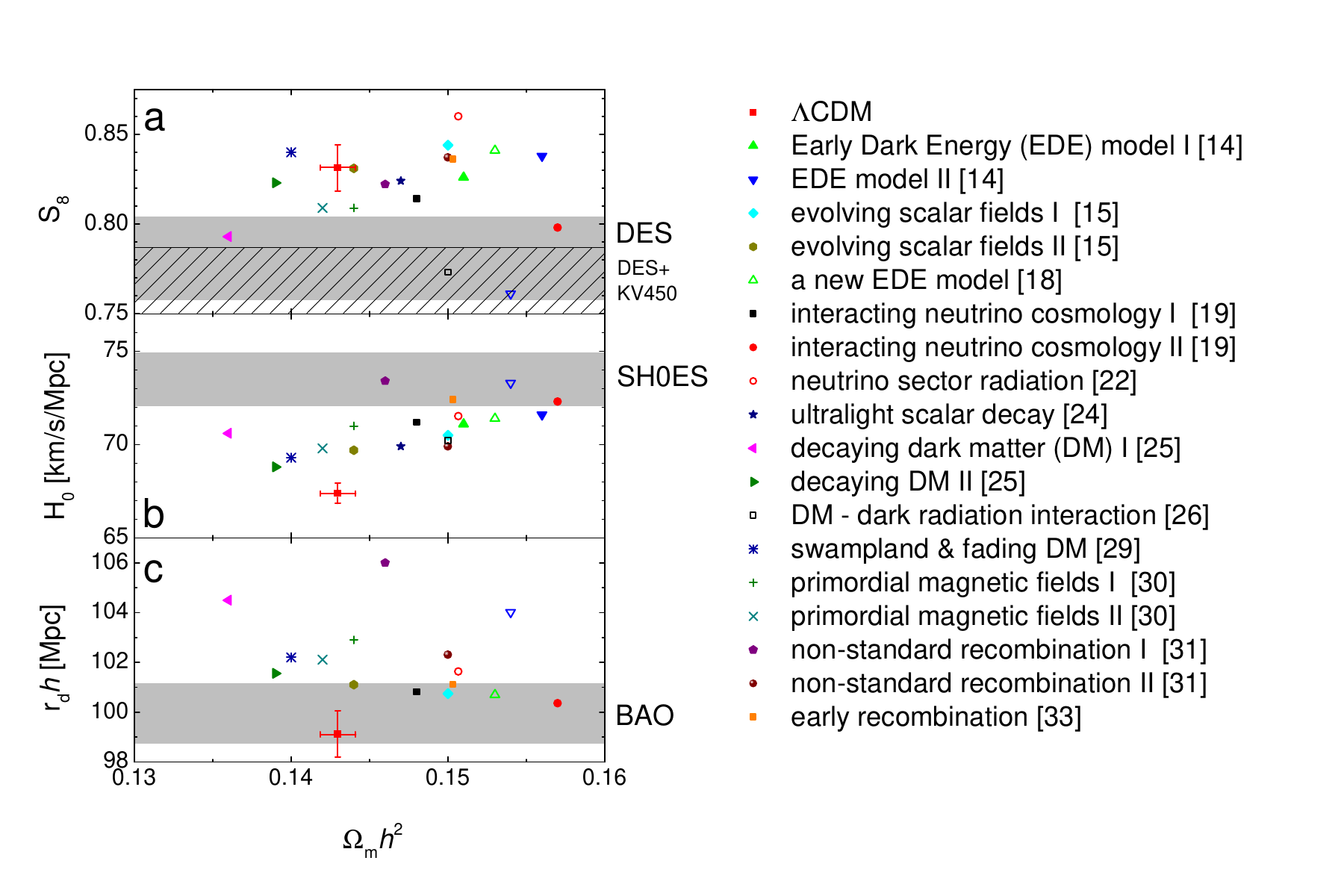}
\caption{\textit{Left panel}: degeneracy lines in the $r_d$-$H_0$ plane defined by constant $\theta_{\star}$ and $\theta_d$ (the latter at different redshifts), and for different values of $\Omega_mh^2$ in the case of $\theta_{\star}$, alongside constraints from \textit{Planck} CMB data (within $\Lambda$CDM), then-current BAO data, and the SH0ES $H_0$ measurement. \textit{Right panel}: best-fit values of $\Omega_mh^2$, $r_d$, $H_0$, and $S_8$ for a compilation of models aiming to solve the Hubble tension, alongside 68\% constraints on the corresponding parameters from different surveys (grey bands). No single model manages to solve the Hubble tension while remaining consistent with all other external measurements. Reproduced from Fig.~2 and Fig.~3 of Jedamzik, Pogosian \& Zhao, ``\textit{Why reducing the cosmic sound horizon alone can not fully resolve the Hubble tension}'', \textit{Communications Physics}, Volume 4, article number 123, \href{https://www.nature.com/articles/s42005-021-00628-x}{doi:10.1038/s42005-021-00628-x}, published 8 June 2021~\cite{Jedamzik:2020zmd}, with permission from the authors. Both figures are licensed under a \href{https://creativecommons.org/licenses/by/4.0/}{Creative Commons Attribution 4.0 International License}, and no changes were made.}
\label{fig:baowl}
\end{center}
\end{figure*}

On the other hand, BAO measurements at a certain redshift $z_{\rm obs}$ (typically $z_{\rm obs} \lesssim 2.5$) carry the imprint of $r_d$, the sound horizon at baryon drag. The latter is the epoch when baryons were released from the drag of photons, and takes place at a redshift $z_d$ slightly lower than $z_{\star}$, making $r_d$ slightly larger than $r_{\star}$ as a result: in essentially all reasonable modified recombination scenarios, $r_d \approx 1.0184 r_{\star}$ holds. Focusing for simplicity on transverse BAO measurements, these constrain the BAO angular scale $\theta_d(z_{\rm obs})$,~\footnote{Note that in Ref.~\cite{Jedamzik:2020zmd} and in Fig.~\ref{fig:baowl} $\theta_d$ is denoted by $\theta^{\rm BAO}$.} given by the following:
\begin{eqnarray}
\theta_d(z_{\rm obs}) \equiv \frac{r_d}{D(z_{\rm obs})}\,,
\label{eq:thetabao}
\end{eqnarray}
where once again $D(z_{\rm obs})$ is the comoving distance to the (effective, survey-averaged) redshift at which the BAO feature is observed. Observing the BAO feature at several redshifts allows one to constrain $\Omega_m$ and the product $r_dh$, or equivalently $r_{\star}h$, given the assumed relation between $r_d$ and $r_{\star}$. Eqs.~(\ref{eq:thetastarcmb},\ref{eq:thetabao}) together with the assumed $r_d$-$r_{\star}$ relation are the starting point for the work of Jedamzik, Pogosian and Zhao~\cite{Jedamzik:2020zmd}, which investigated the effect of early-time new physics decreasing $r_{\star}$ (and thereby $r_d$), treating the latter as a free parameter to be as model-independent as possible, while assuming that the post-recombination Universe is described by $\Lambda$CDM.

The key observation of Ref.~\cite{Jedamzik:2020zmd} is that, given a value of $\omega_m$, both CMB and BAO measurements define a  degeneracy line in the $r_d$-$H_0$ plane, along which the corresponding $\theta_{\star}$ and $\theta_d$ are constant (it is in this sense that I adopt the qualifier ``degeneracy''). Importantly, given the enormous difference between $z_{\rm obs}$ and $z_{\star} \gg z_{\rm obs}$, the redshifts at which the acoustic feature is observed in BAO and CMB measurements, the corresponding degeneracy slopes are very different, as shown in the left panel of Fig.~\ref{fig:baowl}. Placing $r_d$ and $H_0$ on the horizontal and vertical axes respectively, the $r_d$-$H_0$ degeneracy lines get steeper with increasing observation redshift, and are therefore steepest when considering the CMB measurements of $\theta_{\star}$. Along the same plane, as shown in the left panel of Fig.~\ref{fig:baowl}, increasing $\omega_m$ moves the degeneracy lines towards the left, keeping the degeneracy slope fixed.

The difference between the BAO and CMB $r_d$-$H_0$ degeneracy slopes plays a key role in assessing the ability to solve the Hubble tension of new physics which \textit{only} lowers $r_d$. To see this, consider again the left panel of Fig.~\ref{fig:baowl}, where the solid red line has been plotted assuming the best-fit value of $\omega_m=0.143$ determined by the \textit{Planck} satellite (whose constraints in the $r_d$-$H_0$ plane are given by the green contours, and are in agreement with BAO data -- purple bands -- for low values of $H_0$). Clearly, for \textit{Planck} and SH0ES measurements (grey bands) to be brought into agreement, one has to move along one of the $r_d$-$H_0$ degeneracy lines: if this is done along the red line (``low'' $\omega_m$, \textit{Planck} best-fit), one clearly sees that it is impossible to obtain complete consistency between CMB, SH0ES, \textit{and} BAO data, given that the point where the red line and the grey bands meet is outside the purple bands. 

The only way for a CMB $r_d$-$H_0$ degeneracy line to meet both BAO and SH0ES measurements is for this line to be obtained at a higher value of $\omega_m$: see for instance the green dotted line (in the example of Fig.~\ref{fig:baowl}, ``high'' $\omega_m=0.167$). However, in this minimal scenario where the effect of new physics is \textit{only} that of reducing $r_d$, the increase in $\omega_m$ quickly bring a tension with WL data, worsening the already existing $S_8$ discrepancy. Recall that, along with $\Omega_m$, the parameter best constrained by WL measurements is $S_8 \equiv \sigma_8\sqrt{\Omega_m/0.3}$, where $\sigma_8$ is the present day linear theory amplitude of matter fluctuations averaged in spheres of radius $8\,h^{-1}{\rm Mpc}$, and quantifies the variance of fluctuations on this scale. The clustering amplitude $S_8$ depends both on the amplitude of the primordial power spectrum (fixed by the CMB), as well as the net growth of matter perturbations, which increases as $\omega_m$ is increased. Therefore raising $\omega_m$, which is required to achieve consistency between CMB, SH0ES, and BAO measurements as argued earlier, comes at the price of increasing $S_8$: this exacerbates the $S_8$ discrepancy, as the values of $S_8$ preferred by WL surveys such as DES and KiDS are already lower than that inferred from \textit{Planck}, with the significance of this discrepancy lying between $2$ and $3\sigma$ depending on the underlying assumptions (see Ref.~\cite{Nunes:2021ipq}).

The overall observation of Jedamzik, Pogosian and Zhao~\cite{Jedamzik:2020zmd} is therefore that early-time new physics which reduces the sound horizon \textit{alone} cannot completely resolve the Hubble tension, as it will always create additional tensions: with BAO data if operating at lower $\omega_m$, and with WL data (worsening the $S_8$ discrepancy) if operating at higher $\omega_m$. These problems were first explicitly noted in the context of early dark energy (EDE) models, which introduce a dark energy-like component dynamically relevant around matter-radiation equality, and where the necessity of operating at higher $\omega_m$ is related to the enhanced early integrated Sachs-Wolfe effect predicted by such models (which will become relevant in discussing my fifth hint in Sec.~\ref{subsec:eisw}). Surveying a compilation of models introduced to solve the Hubble tension, Jedamzik, Pogosian and Zhao~\cite{Jedamzik:2020zmd} confirmed that the features discussed previously are indeed generically observed, in some cases more strongly than others (see the right panel of Fig.~\ref{fig:baowl}).

To conclude, this second hint indicates that early-time new physics whose only effect is to reduce the sound horizon cannot fully solve the Hubble tension. It is worth noting that the vast majority of early-time new physics models introduced in this context were introduced \textit{precisely} to lower the sound horizon: one notable exception is the strongly interacting neutrino model, which however is now severely constrained by CMB polarization data. Clearly, a full solution to the Hubble tension which does not introduce additional tensions requires extra ingredients beyond the $r_d$-lowering ones. While I will discuss in more detail some possibilities in this direction in Sec.~\ref{sec:promising}, here I simply note that including \textit{additional} early-time new physics is highly non-trivial due to the necessity of maintaining a good fit to \textit{all} features in the CMB temperature and polarization spectra, beyond just the acoustic peaks. I believe a much more promising direction involves additional late-time new physics, which ``decouples'' itself from the early-time effects, and possibly predominantly operating at the perturbation level in order not to spoil $\Lambda$CDM's fit to late-time background measurements.

\subsection{Cosmic chronometers}
\label{subsec:cc}

To further arbitrate the Hubble tension or, as in our case, shed light on which epochs may require new physics, it is helpful to consider cosmological datasets which are as model-independent as possible.~\footnote{Here by ``model-independent'' I really mean independent of any assumed underlying \textit{cosmological} model, where the cosmological model-dependence of the ``standard'' CMB, BAO, SNeIa probes may be traded for dependence on other models, e.g.\ of more astrophysical/astronomical nature.} In order to identify whether late-time new physics may be required, it is helpful to consider probes which carry little or no dependence on early-time physics, as with the OAO considered earlier. Should such a probe return a value of $H_0$ consistent with the high-redshift ones (e.g.\ the \textit{Planck} low value of $H_0$ assuming $\Lambda$CDM), the resulting (persisting) tension with the local measurements of $H_0$ is, by construction, one that may not be resolved by early-time new physics.

Cosmic chronometers (CC) are interesting in this sense, as they are able to provide a direct, cosmology-independent estimate of the Hubble rate $H(z)$. They were first proposed by Jim\'{e}nez and Loeb~\cite{Jimenez:2001gg}, and rely on inverting the time-redshift relation within a Friedmann-Lema\^{i}tre-Robertson-Walker (FLRW) Universe:
\begin{eqnarray}
&&\frac{dt}{dz} = -\frac{1}{(1+z)H(z)} \nonumber \\ \implies
&&H(z) = -\frac{1}{1+z}\frac{dz}{dt}=-\frac{1}{1+z} \left ( \frac{dt}{dz} \right ) ^{-1} \,,
\label{eq:hzcc}
\end{eqnarray}
valid for a FLRW Universe and assuming only homogeneity, isotropy, and a metric theory of gravity, with no further assumptions on the functional form of the expansion rate or the spatial geometry. Using Eq.~(\ref{eq:hzcc}) requires identifying a class of ``cosmic chronometers'' to estimate $dt/dz$ as the look-back time differential change with redshift: as redshifts (and hence $dz$) can easily be measured via spectroscopy, the difficulty is in finding a trustworthy estimator for look-back time (and hence $dt$).

The question of which objects are best suited for use as CC has been the subject of much study over the past two decades. It is now understood that extremely massive ($M \gtrsim 10^{11}M_{\odot}$), early, passively-evolving galaxies (i.e.\ evolving on a timescale much larger than their differential ages) are an excellent choice~\cite{Thomas:2009wg}. The reason is that these galaxies formed and assembled their mass at high redshift ($z \sim 2$-$3$) and over a very short period of time ($t \lesssim 0.3\,{\rm Gyr}$), after which they quickly exhaust their gas reservoir and evolve passively. With a suitable CC sample at hand, the age difference $\Delta t$ between two passively-evolving CC formed approximately at the same time and separated by a small redshift interval $\Delta z$ around $z_{\rm eff}$ can be used to obtain $dz/dt \approx \Delta z/\Delta t$, and thus $H(z_{\rm eff})$ via Eq.~(\ref{eq:hzcc}). While not unrelated to the OAO discussed in Sec.~\ref{subsec:oao}, the crucial difference is that CC rely on \textit{differential} rather than absolute age measurements, and hence are affected to a much lesser degree by systematics, which are expected to (at least partially) cancel when considering age differences.

The appeal of CC lies in their providing a determination of $H(z)$ free from cosmological model assumptions. To infer $H_0$ from CC data, one can proceed non-parametrically (e.g.\ as done by Yu, Ratra and Wang~\cite{Yu:2017iju}, G\'{o}mez-Valent and Amendola~\cite{Gomez-Valent:2018hwc}, and many other works), or assuming a specific model. Two important points are worth noting now:
\begin{itemize}
\item regardless of whether one goes parametric or non-parametric, the value of $H_0$ inferred from CC is the \textit{cosmological} (as opposed to local) one;
\item as CC directly measure the late-time ($z \lesssim 2$) expansion rate, a parametric analysis thereof requires no assumption whatsoever about early Universe physics (this is actually true for a non-parametric inference as well).
\end{itemize}
Focusing on the parametric approach, which is the one we shall follow later, two immediate corollaries of the above are:
\begin{itemize}
\item should the value of $H_0$ inferred from CC within a certain model be in tension with local $H_0$ measurements, such a conclusion would be \textit{completely independent} of whatever happened in the early Universe, including before and around recombination;
\item assuming such a (residual) tension is physical (i.e.\ not due to systematics), resolving it would require introducing new (relative to the assumed parametric model) late-time ingredients to alter the cosmological value of $H_0$, and/or new local physics to alter the local value of $H_0$.
\end{itemize}
The above points suggest that CC can be used as a late-time consistency test of $\Lambda$CDM, much as the OAO discussed in Sec.~\ref{subsec:oao}, by comparing the value of $H_0$ inferred from CC assuming $\Lambda$CDM to local $H_0$ measurements, and assessing whether there is a (residual) tension between the two. Of course, CC data come with several caveats pertaining to their reliability similar to those of OAO (although less severe, given the fact that the CC galaxies are passively evolving), and inevitably carry dependence on assumed astrophysical galaxy evolution models. The fact that CC are differential rather than absolute age measurements somewhat mitigates these potential systematics, which have nevertheless been studied and quantified in detail in various works, including several by Moresco \textit{et al.}~\cite{Moresco:2018xdr,Moresco:2020fbm,Moresco:2022phi}: therefore, at the current stage of things, CC data are significantly more trustworthy than OAO.

I perform this analysis on the latest compilation of 32 CC measurements, comprising the 31 measurements listed in Tab.~1 of Ref.~\cite{Vagnozzi:2020dfn} and the latest measurement of Borghi \textit{et al.}~\cite{Borghi:2021rft}, including systematics and off-diagonal covariance terms as discussed in detail in various recent works (e.g.\ Ref.~\cite{Moresco:2022phi}).~\footnote{I follow the procedure outlined in \href{https://gitlab.com/mmoresco/CCcovariance}{gitlab.com/mmoresco/CCcovariance}.} Assuming $\Lambda$CDM, I infer $H_0=67.8 \pm 3.0\,{\rm km}/{\rm s}/{\rm Mpc}$ and $\Omega_m=0.33 \pm 0.06$: both values are in excellent agreement with the values inferred from the \textit{Planck} satellite assuming $\Lambda$CDM, although the uncertainties are a factor of respectively $6$ and $8$ bigger, which should not come as a surprise, given that I have analyzed CC data on their own. One could of course try and sharpen these inferences using additional external datasets (e.g.\ a prior on $\Omega_m$ from weak lensing or cluster count measurements), but here I choose to be as conservative as possible and use no data other than CC, in order to keep the picture as clean as possible.

\begin{figure*}
\begin{center}
\includegraphics[width=0.7\linewidth]{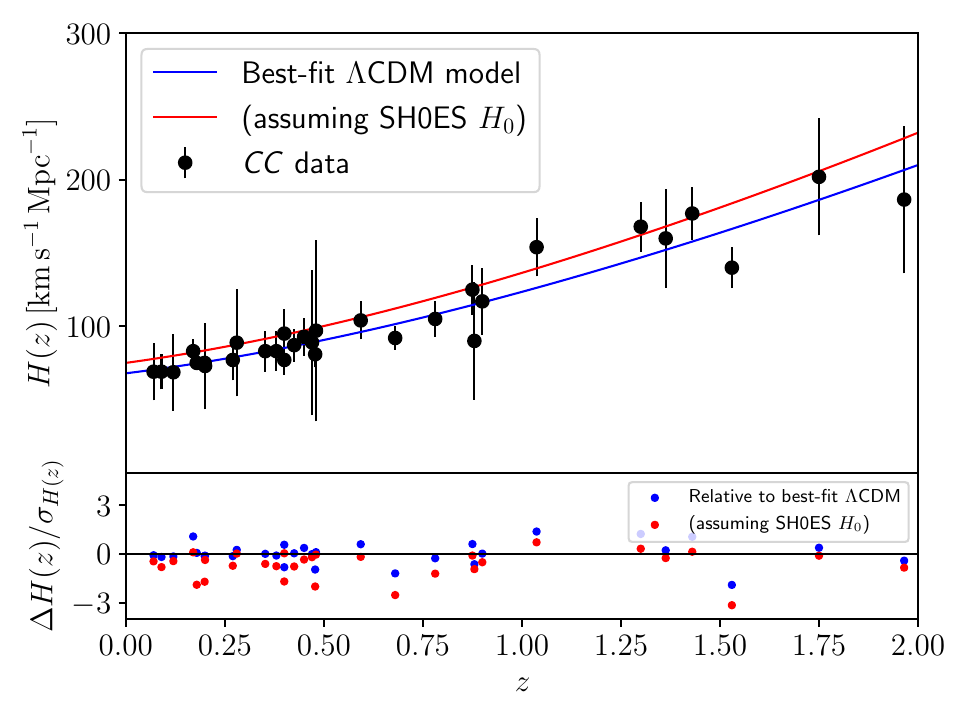}
\caption{\textit{Upper panel}: CC measurements (black datapoints), alongside the predicted expansion rate given the best-fit $\Lambda$CDM cosmological parameters from a fit to CC data alone (blue curve), and the same cosmology but with $H_0$ fixed to the SH0ES value. \textit{Lower panel}: residuals (CC data minus theory predictions) in units of datapoint uncertainties, for the two cosmologies considered in the upper panel, with the same color coding.}
\label{fig:cc}
\end{center}
\end{figure*}

While the uncertainty on $H_0$ is relatively large, it is intriguing that the central value is in remarkable agreement with the \textit{Planck}'s central value ($67.4\,{\rm km}/{\rm s}/{\rm Mpc}$). Moreover, despite the size of the uncertainty, a $\lesssim 2\sigma$ tension with the local $H_0$ determination from Cepheid-calibrated SNeIa remains. As with the case of OAO earlier, this residual tension is not high enough to be alarming, but nonetheless worthy of attention. It is worth examining which part(s) of the CC dataset prevent a high $H_0$ determination. To show this, in Fig.~\ref{fig:cc} I plot the CC dataset alongside the $\Lambda$CDM prediction for the expansion rate (upper panel blue curve), with the best-fit values for the parameters obtained by analyzing the CC dataset alone, and with the corresponding residuals (data minus theory, in units of data uncertainties) shown in the lower panel. Then, I plot the same functional form for the expansion rate, but with $H_0$ fixed to the local SH0ES value (red curve). From the lower panel (red points), it is clear that there is no single point driving the preference for low $H_0$ (or rather disfavoring high $H_0$). Rather, there are a number of points (especially for $z \lesssim 1$, as well as a point around $z \approx 1.5$) for which the high $H_0$ fit is noticeably worse than the low $H_0$ one, with the high $H_0$ theory prediction in some cases being nearly $3\sigma$ off from the data.

If we therefore take this residual tension seriously, and eliminate the possibility of systematics, once more we are left with the options of questioning the validity of $\Lambda$CDM at late times, and/or the validity of local $H_0$ determination(s). Should one pursue the first route, new late-time physics should again go in the direction of lowering the expansion rate relative to $\Lambda$CDM at $z>0$ (similar to the effect of a phantom component), compatibly with BAO and high-$z$ SNeIa constraints. As with the earlier discussion on OAO, I once more note that the possibilities of new late-time (cosmological) and local physics are not mutually exclusive: the two can act simultaneously (together with early-time new physics), the former to raise the cosmological $H_0$ value, the latter to lower the local $H_0$ value, to bring the two into better agreement and meet along the way. In closing, it is intriguing that both absolute (Sec.~\ref{subsec:oao}) and relative (Sec.~\ref{subsec:cc}) galaxy ages independently~\footnote{The OAO and CC datasets are mostly independent, except for 8 CC measurements with relatively large uncertainties~\cite{Simon:2004tf}, whose relative statistical weight in the conclusions is therefore low. Note that the reliability of these measurements has also been recently questioned by Kjerrgren and M\"{o}rtsell~\cite{Kjerrgren:2021zuo}. Given their low statistical weight in reaching my conclusions, a posteriori this should not present a concern.} appear to disfavor high ($\gtrsim 73\,{\rm km}/{\rm s}/{\rm Mpc}$) values of $H_0$ and, when used as consistency tests of $\Lambda$CDM, indicate the need for new physics at late times and/or local scales, if the residual tension with local $H_0$ measurements is taken seriously.

\subsection{Descending trends in low-redshift data}
\label{subsec:trends}

The $\Lambda$CDM model is a dynamical model (i.e.\ one which describes the evolution of a system over time or, equivalently, redshift), equipped with a number of fitting parameters. In science, including physics and by extension cosmology, it is a non-negotiable (mathematical) fact that dynamical models break down when fitting parameters which are supposed to be constant actually evolve with time -- more precisely, when their \textit{inferred} values evolve depending on the time when they are inferred. In the case of $\Lambda$CDM, this would correspond to the inferred values of otherwise constant parameters taking different values depending on the redshift of the data used to infer the parameters themselves. Let me now be more specific and focus on $H_0$ as the parameter of interest. Within an FLRW Universe, from the mathematical point of view $H_0$ is none other than an integration constant: therefore, by definition, it should be constant regardless of the redshift at which it is inferred. More specifically, as shown by Krishnan \textit{et al.}~\cite{Krishnan:2020vaf}, under the assumptions of homogeneity and isotropy, and once a theoretical in the form of an effective equation of state $w_{\rm eff}(z)$,~\footnote{This effective equation of state directly enters the second Friedmann equation (the acceleration equation) and includes contributions from \textit{all} species, not only dark energy.} the Friedmann equations can be integrated to give:
\begin{eqnarray}
H_0 = H(z)\exp \left [ -\frac{3}{2}\int_0^zdz'\,\frac{1+w_{\rm eff}(z')}{1+z'} \right ] \,.
\label{eq:runninghubbletension}
\end{eqnarray}
While Eq.~(\ref{eq:runninghubbletension}) is a mathematical identity, from the observational point of view an useful interpretation is as follows. An input $w_{\rm eff}(z)$, appearing in the right-hand side of Eq.~(\ref{eq:runninghubbletension}) is specified in the form of a underlying model fit to the data, the latter corresponding to $H(z)$ appearing on the same side of the equation.~\footnote{Of course the data need not be exactly in the form of $H(z)$, but other types of data (e.g.\ distance measurements) can be brought into this form, or conversely Eq.~(\ref{eq:runninghubbletension}) can be generalized to account for other types of data.} This is used to infer $H_0$ on the left-hand side of the equation. If the input [$w_{\rm eff}(z)$] and the data [$H(z)$] ``agree'', the value of $H_0$ inferred as a function of the redshift of the data should be consistent (within uncertainties) across redshift. On the other hand, if $w_{\rm eff}(z)$ and $H(z)$ disagree beyond the uncertainties, $H_0$ picks up $z$ dependence and ``runs''. Note that this statement can also be made for other parameters beyond $H_0$ (e.g.\ $\Omega_m$ and $S_8$) which are either integration constants, or are directly related to integration constants.

Overall, it would appear that inferring running constants is a telltale signature of a model's death. In some way, the $H_0$ tension (and other tensions) can be thought of as a primitive example of such running, in the simplest case between two (extreme) redshifts. However, if the $H_0$ tension is physical (i.e.\ not due to systematics) \textit{and} calls for some amount of late-time new physics,~\footnote{In principle, if some early-time new physics raised the CMB value of $H_0$ to be perfectly in agreement with the local value, and the late-time Universe were completely described by $\Lambda$CDM, there should be no evolving trend at intermediate redshifts.} redshift evolution at some other intermediate redshift is non-negotiable. If such a redshift evolution is not observed, one is left with the conclusion that either systematics are to blame, and/or that the assumption of FLRW itself [underlying the derivation of Eq.~(\ref{eq:runninghubbletension})] should be dropped. This is a mathematical statement, and amounts to the observation that there cannot be contradictions between mathematics and observations. This leads to the following questions:
\begin{itemize}
\item Has such a redshift evolution already been observed in current data?
\item Has it been observed across different independent datasets, and if so is there a common trend across these independent datasets?
\item Are the size and/or direction of this trend inconsistent with what one would expect if no new physics were at play, i.e.\ can it be attributed to new physics?
\end{itemize}
With a few caveats, the answers to the above three questions are positive: hints of running $H_0$ have now been observed in multiple datasets, and in all of these $H_0$ decreases with increasing redshift, which is why I will refer to these as ``descending trends''.
\begin{figure*}
\begin{center}
\includegraphics[width=0.7\linewidth]{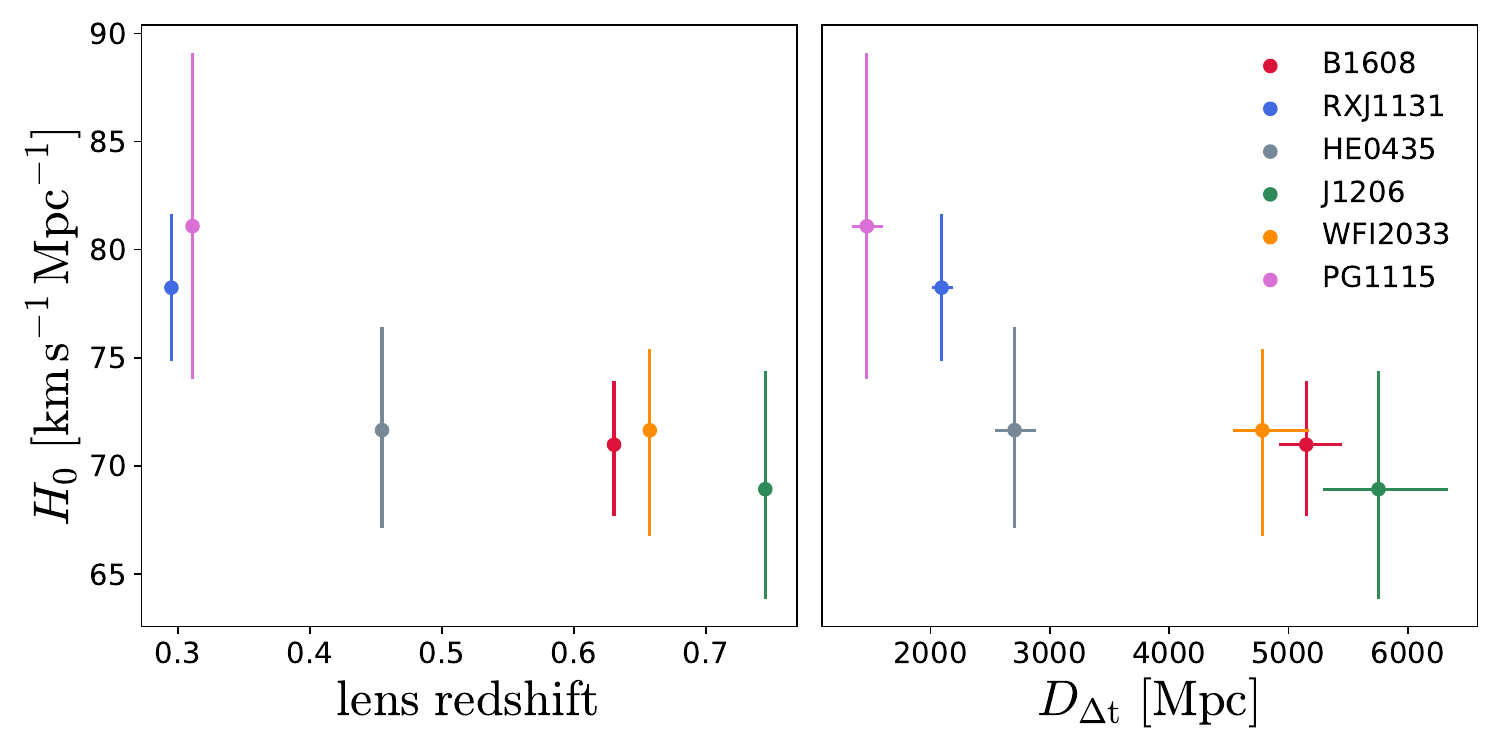}
\caption{\textit{Left panel}: constraints on $H_0$ from the individual H0LiCOW lenses as a function of lens redshift. \textit{Right panel}: same as for the left panel, but as a function of the lens time-delay distances. Reproduced from Fig.~A1 of Wong \textit{et al.}, ``\textit{H0LiCOW -- XIII. A 2.4 per cent measurement of $H_0$ from lensed quasars: 5.3$\sigma$ tension between early- and late-Universe probes}'', \textit{Monthly Notices of the Royal Astronomical Society}, Volume 498, Issue 1, Pages 1420-1439, \href{https://academic.oup.com/mnras/article/498/1/1420/5849454}{doi:10.1093/mnras/stz3094}, published 16 September 2019~\cite{Wong:2019kwg}. \textcopyright\,(2019) The Authors. Reproduced by permission of Oxford University Press on behalf of the Royal Astronomical Society. All rights reserved.}
\label{fig:h0licowdescendingtrend}
\end{center}
\end{figure*}

The first, and perhaps most famous, example of descending trend was reported by the H0LiCOW collaboration, who in Wong \textit{et al.}~\cite{Wong:2019kwg} performed a joint analysis of six gravitationally lensed quasars with measured time delays to infer $H_0$. The strong lensing approach is a one-step technique, which unlike the distance ladder approach requires no external calibration, although the absence of a ladder is traded for the dependence on assumptions concerning the lens and line-of-sight mass distribution~\cite{Birrer:2015fsm,Kochanek:2019ruu,Blum:2020mgu}. As such, the method is completely independent of and complementary to the CMB and distance ladder approaches. Nevertheless, the inferred value of $H_0$ is still cosmological model-dependent, as it is related to the so-called ``time-delay distance'' $D_{\Delta t}$. The time-delay distance combines information on the lens redshift as well as the angular diameter distances to the lens, source, and between lens and source. Therefore, interpreting $D_{\Delta t}$ measurements requires adopting an underlying cosmological model, which in the case of H0LiCOW was chosen to be the $\Lambda$CDM model.

Albeit relegated to an Appendix, the H0LiCOW collaboration~\cite{Wong:2019kwg} noted a curious trend where the value of $H_0$ inferred from each of the single lenses decreased with increasing redshift lens. This trend, in the redshift range $0.3 \lesssim z \lesssim 0.7$, is reported in the left panel of Fig.~\ref{fig:h0licowdescendingtrend}, whereas the right panel reports a similar trend observed with the time-delay distance. In particular, the central value of $H_0$ evolves from $\sim 82\,{\rm km}/{\rm s}/{\rm Mpc}$ when inferred from the lens PG1115 at redshift $z \sim 0.3$, to $\sim 68\,{\rm km}/{\rm s}/{\rm Mpc}$ when inferred from the lens J1206 at redshift $\sim 0.7$. Using mock data, the $H_0$ trend was found to deviate from the null hypothesis (where $H_0$ does not run with redshift) at an equivalent Gaussian significance level of nearly $2\sigma$~\cite{Wong:2019kwg}. Albeit intriguing, the trend and its possible physical origin were not discussed further, possibly also due to its moderate significance. A similar trend was confirmed by the TDCOSMO collaboration when revisiting the analysis including stellar kinematics measurements to break the mass-sheet degeneracy and thereby jointly infer $H_0$ and lens density profiles~\cite{Birrer:2020tax}. The key point, however, is that this trend is inferred without making any assumptions on early-Universe physics, since the interpretation of the time-delay distance measurements is essentially only sensitive to late-time physics (and, in this case, the assumption of $\Lambda$CDM). Therefore, if not due to systematics or astrophysical mismodeling, ``fixing'' this $2\sigma$ trend~\footnote{At this point there is no reason not to refer to this trend as a $2\sigma$ tension where, if one wants, the tension is with the mathematical requirement that $H_0$ be a (integration) constant.} can only be achieved by fiddling around with late-time new physics.

\begin{figure*}
\begin{center}
\includegraphics[height=5.5cm]{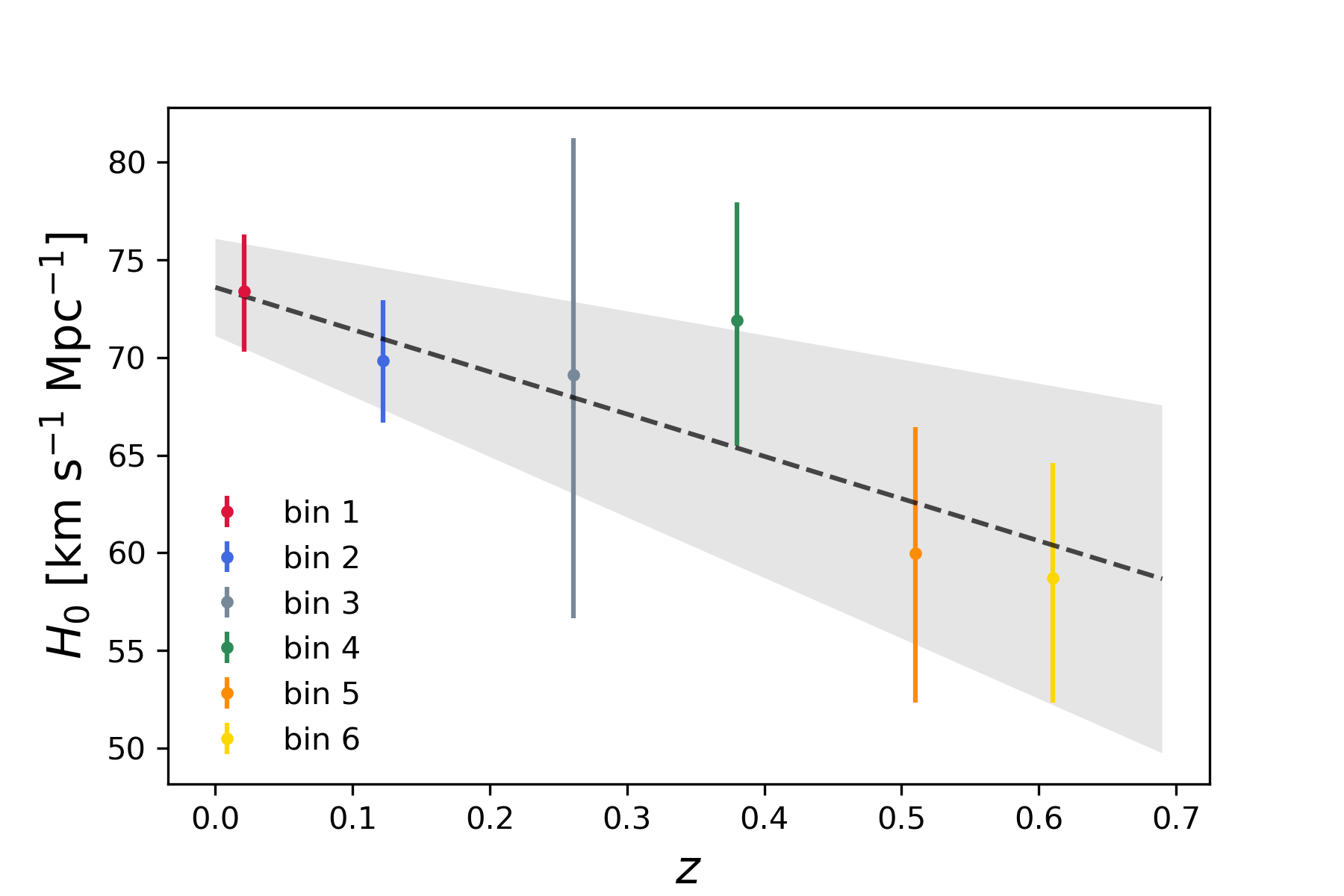}\,\includegraphics[height=5.5cm]{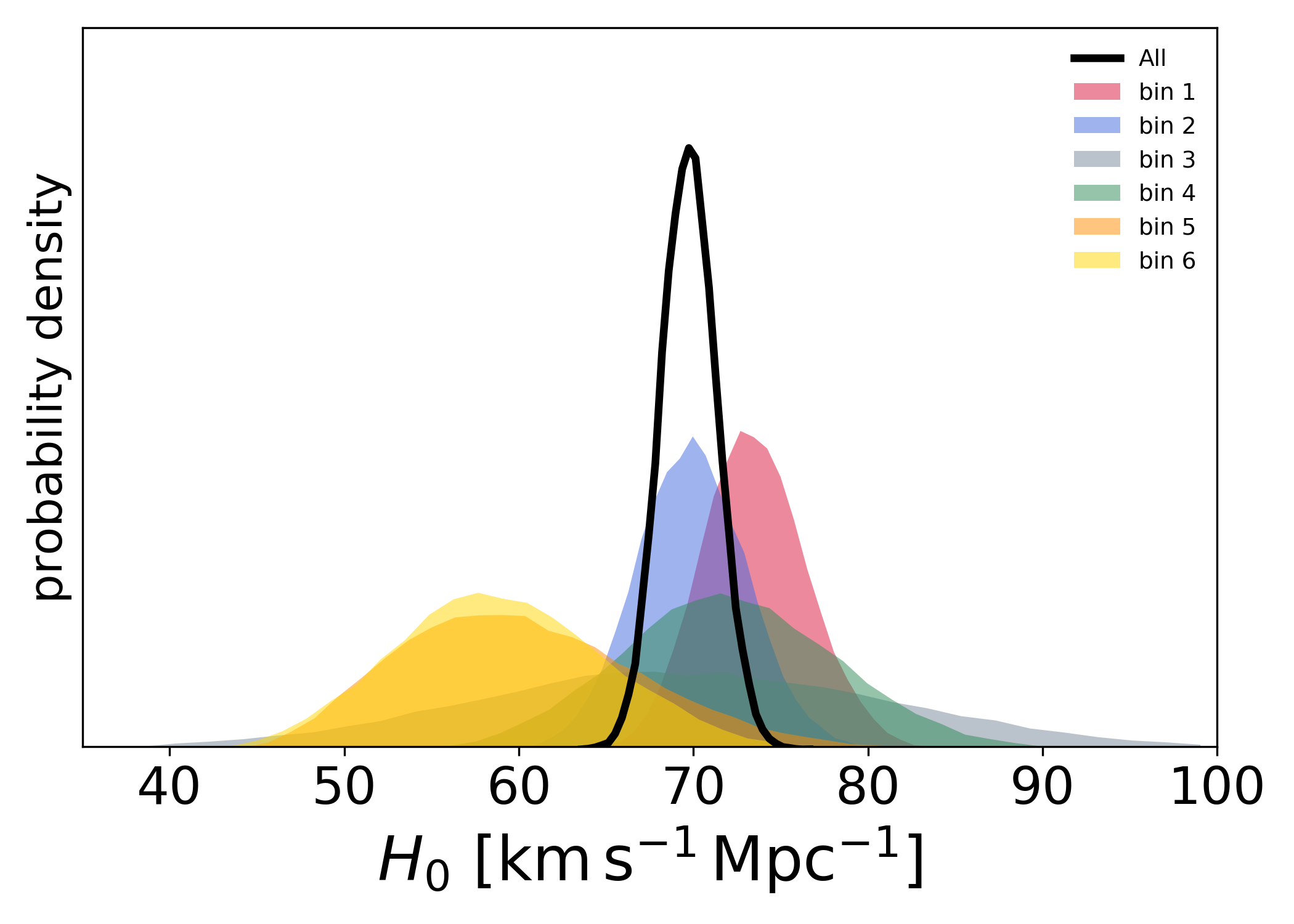}
\caption{\textit{Left panel}: binned constraints on $H_0$ obtained from the late-time dataset combination considered in Ref.~\cite{Krishnan:2020obg}. \textit{Right panel}: probability distribution functions for $H_0$ as inferred from each bin, together with the overall constraint obtained from the complete dataset. Reproduced from Fig.~1 and Fig.~2 of Krishnan \textit{et al.}, ``\textit{Is there an early Universe solution to Hubble tension?}'', \textit{Physical Review D}, Volume 102, Issue 10, article number 103525, \href{https://journals.aps.org/prd/abstract/10.1103/PhysRevD.102.103525}{doi:10.1103/PhysRevD.102.103525}, published 20 November 2020~\cite{Krishnan:2020obg}. \textcopyright\,(2020) American Physical Society. Reproduced by permission of the American Physical Society and the authors. All rights reserved.}
\label{fig:baoccsneiadescendingtrend}
\end{center}
\end{figure*}

Inspired by the H0LiCOW results, Krishnan \textit{et al.}~\cite{Krishnan:2020obg} considered a low-redshift ($z \lesssim 0.7$) dataset which overall indicates an intermediate ($\sim 70\,{\rm km}/{\rm s}/{\rm Mpc}$) value of $H_0$, and used previously by Dutta \textit{et al.}~\cite{Dutta:2019pio}, to examine whether this intermediate value actually hides a similar trend once the data is divided in redshift bins. The adopted dataset included distances to 6 megamaser-hosting galaxies~\cite{Reid:2019tiq,Pesce:2020ghw,Pesce:2020xfe}, CC data restricted to the range $z \leq 0.7$, BAO data from 6dFGS, SDSS-MGS, and BOSS DR12 (only isotropic measurements, without including $f\sigma_8$ measurements), and Pantheon SNeIa in the range $0.01 \leq z \leq 0.7$. Krishnan \textit{et al.}~\cite{Krishnan:2020obg} then divided the data into 6 redshift bins, and inferred $H_0$ from each bin via an MCMC where a number of additional nuisance parameters were also varied (see Ref.~\cite{Krishnan:2020obg} for more details). Note that the presence of CC data, which directly constrain $H(z)$, allows for the use of BAO data without having to impose any external prior on the sound horizon at baryon drag $r_d$, which is instead treated as a free parameter. Therefore, the determinations of $H_0$ are free of any early-Universe physics assumption, and the same is true for any trend that may be inferred. The result of this analysis is shown in Fig.~\ref{fig:baoccsneiadescendingtrend}, where a descending trend broadly similar to the H0LiCOW one is clearly seen. Intriguingly, the slope of the two trends are consistent, albeit the two intercepts differ slightly. Using the same approach as H0LiCOW to establish the significance of the trend, this was found to deviate from the null hypothesis at an equivalent Gaussian significance level of $2.1\sigma$~\cite{Krishnan:2020obg}. As with the H0LiCOW trend, this is by construction one which cannot be fixed by invoking early-time new physics which alters $r_d$, since such modifications will only push the trend up and down without removing it. In fact, Krishnan \textit{et al.}~\cite{Krishnan:2020obg} used this observation to argue that the Hubble tension calls for something more than just early-time new physics.

\begin{figure*}
\begin{center}
\includegraphics[scale=0.10]{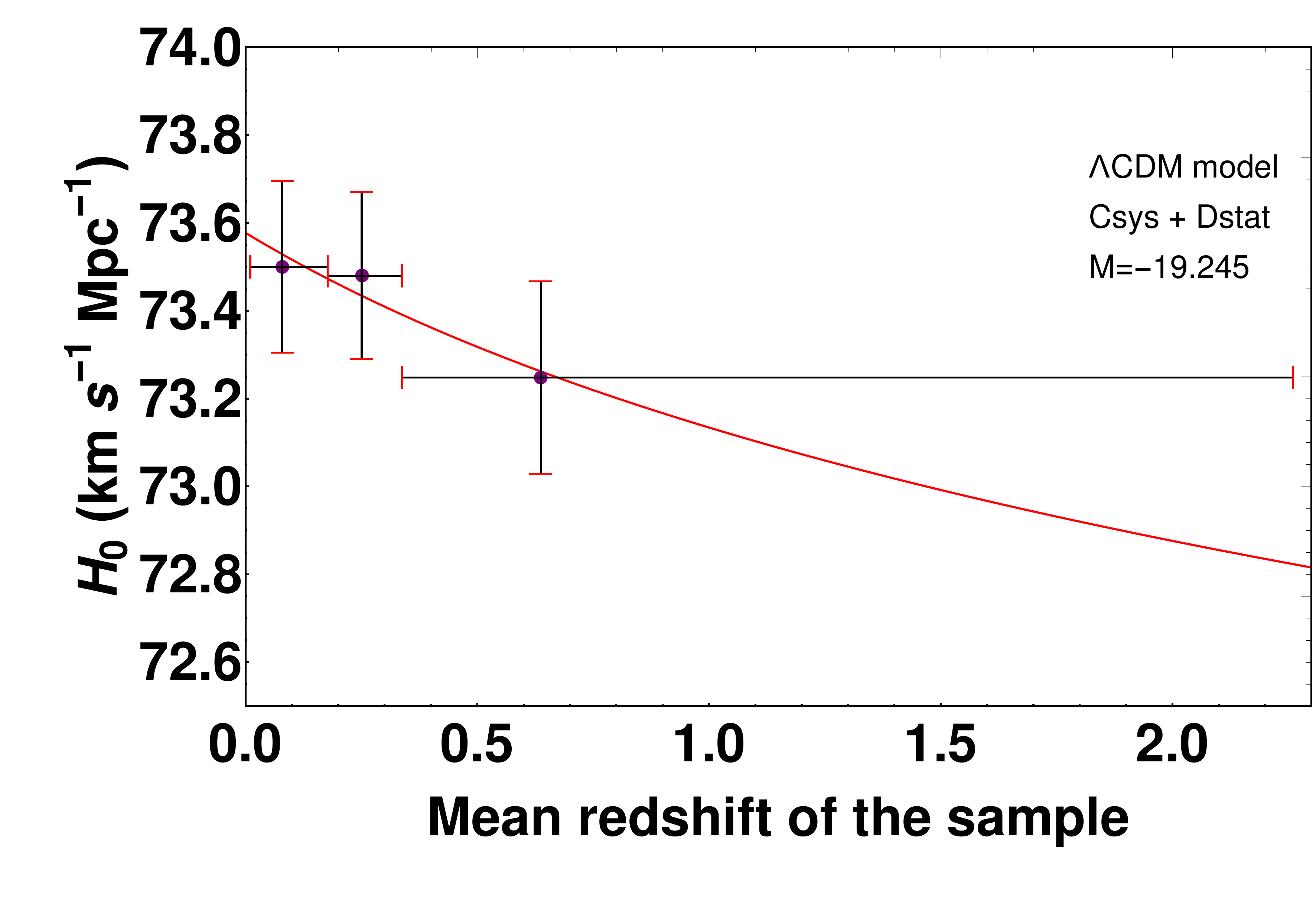}\,\includegraphics[scale=0.10]{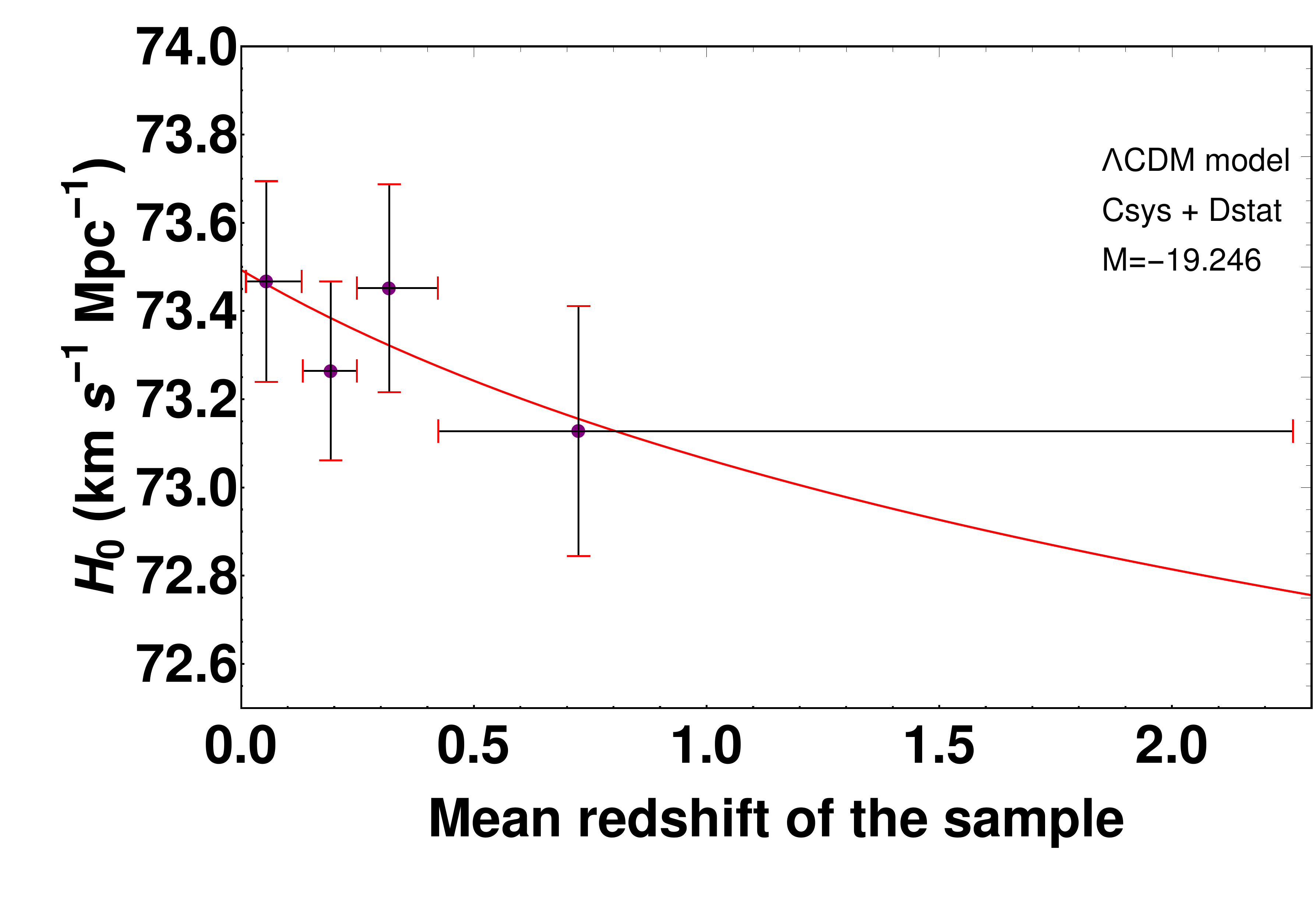}\\
\includegraphics[scale=0.10]{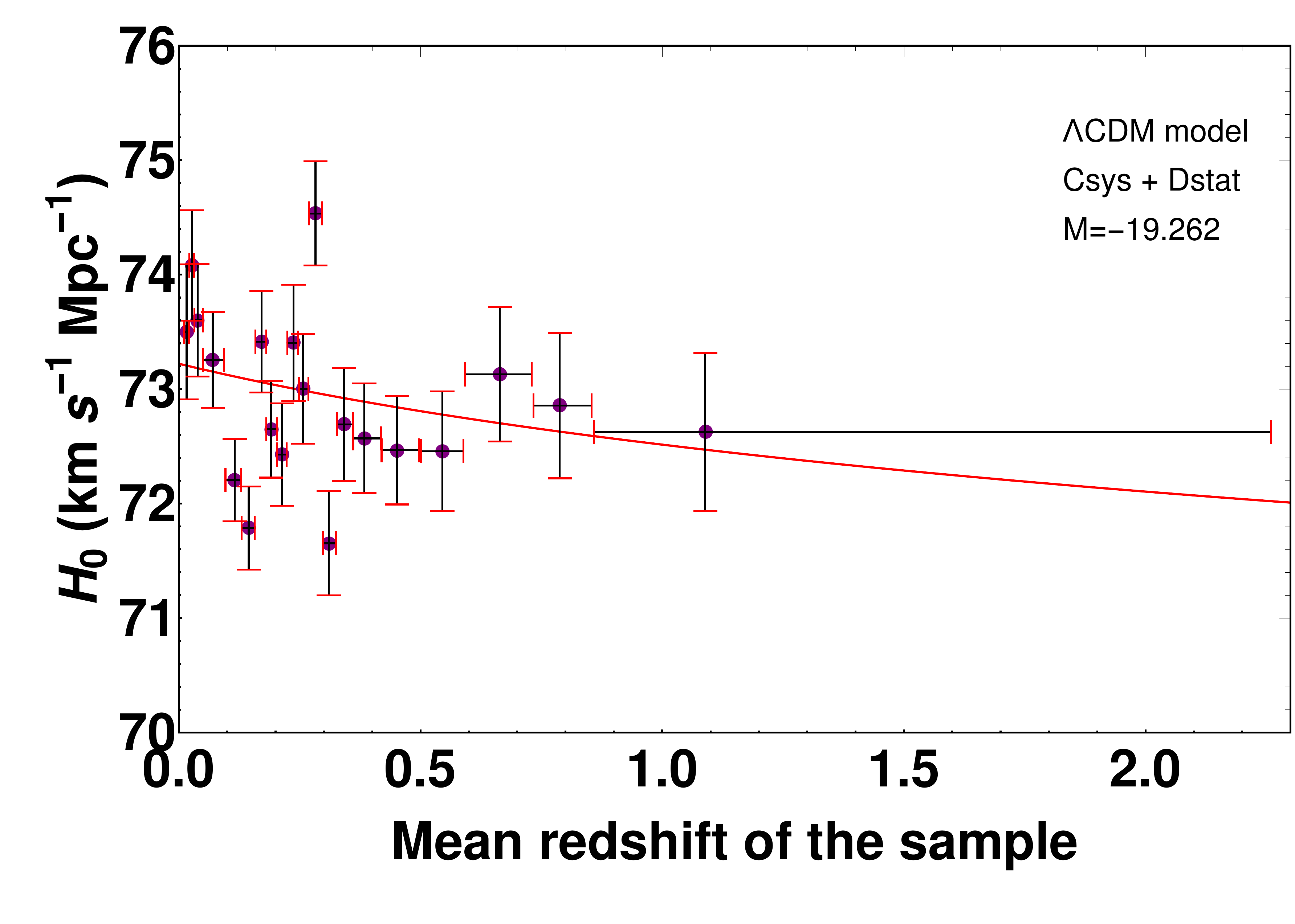}\,\includegraphics[scale=0.10]{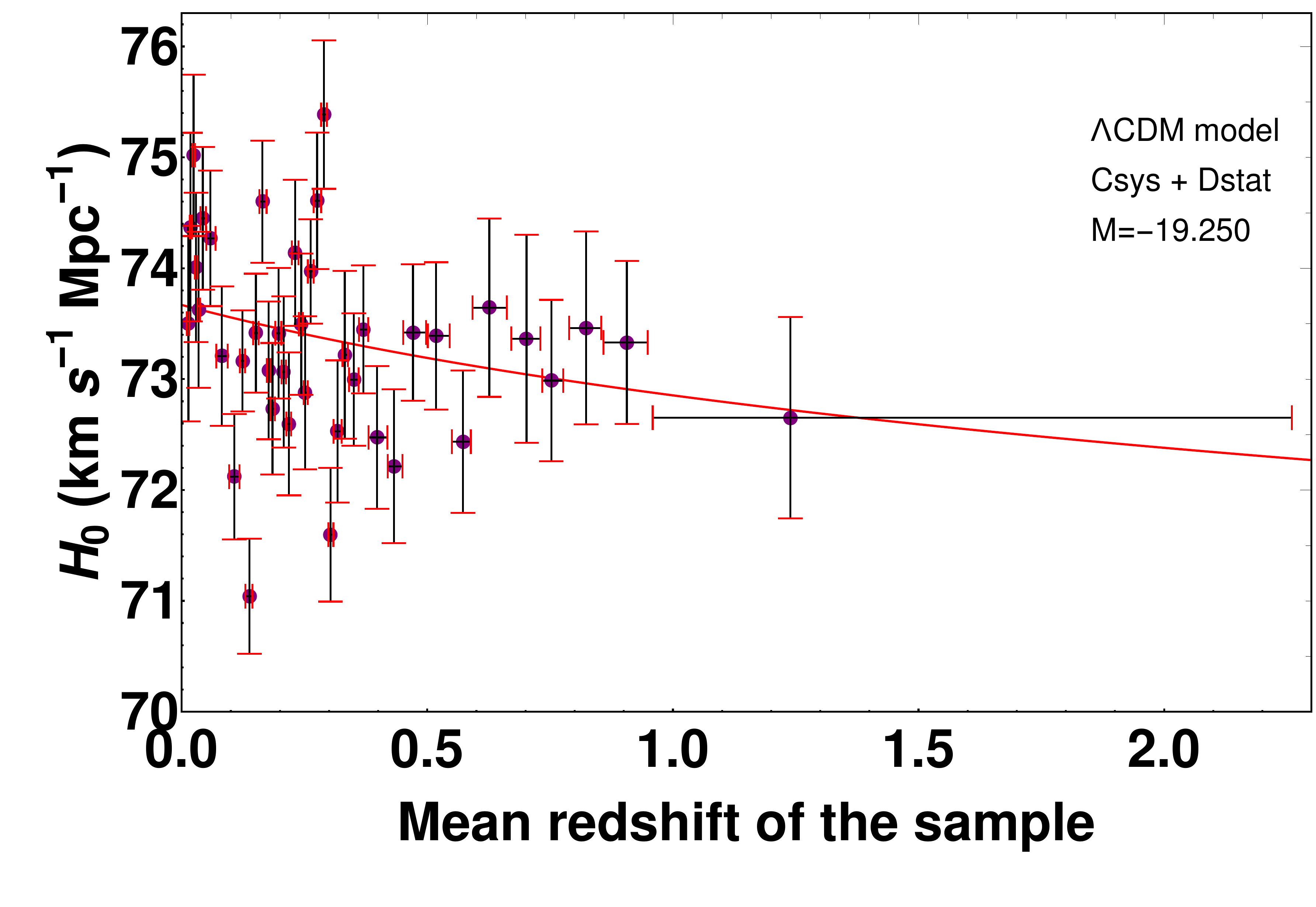}
\caption{Binned constraints on $H_0$ obtained from the \textit{Pantheon} SNeIa sample in Ref.~\cite{Dainotti:2021pqg}, considering 3 redshift bins (upper left panel), 4 redshift bins (upper right panel), 20 redshift bins (lower left panel), and 40 redshift bins (lower right panel). Reproduced from Fig.~5 and Fig.~6 of Dainotti \textit{et al.}, ``\textit{On the Hubble Constant Tension in the SNe Ia Pantheon Sample}'', \textit{The Astrophysical Journal}, Volume 912, Number 2, article number 150, \href{https://iopscience.iop.org/article/10.3847/1538-4357/abeb73}{doi:10.3847/1538-4357/abeb73}, published 17 May 2021~\cite{Dainotti:2021pqg}. \textcopyright\,(2021) American Astronomical Society. Reproduced by permission of the American Astronomical Society and the authors. All rights reserved.}
\label{fig:snei1descendingtrend}
\end{center}
\end{figure*}

Another notable example, and the last one I will discuss to a similar level of detail as the previous two, is the Pantheon SNeIa sample. The analysis of Dainotti \textit{et al.}~\cite{Dainotti:2021pqg} considered the full Pantheon sample and binned the SNeIa into respectively 3, 4, 20, and 40 redshift bins, each containing the same number of SNeIa. Within each bin a value for $H_0$ was then inferred, both assuming the $\Lambda$CDM model, as well as the $w_0w_a$CDM model where the DE equation of state (EoS) is described by the so-called Chevallier-Polarski-Linder parametrization $w(z)=w_0+w_az/(1+z)$. The results, as far as the $\Lambda$CDM model is concerned, are shown in Fig.~\ref{fig:snei1descendingtrend}, with each panel corresponding to a different choice of number of bins. Albeit visually less clear than the previous two, a trend is visible in this case as well, with significance estimated to be up to $2\sigma$ by Dainotti \textit{et al.}~\cite{Dainotti:2021pqg}. When instead assuming the $w_0w_a$CDM model, the trend was found to persist, albeit at a lower significance. The authors then fit the inferred values of $H_0$ with a function of the form $H_0(z)=\widetilde{H}_0(1+z)^{-\alpha}$, where $\widetilde{H}_0=H_0(z=0)$, inferring values of $\alpha$ differing from $0$ at up to $2\sigma$. Various possible astrophysical causes for such a redshift evolution of $H_0$ were discussed (in fact, the proposed fitting function is routinely used in modelling astrophysical evolution effects in Gamma-Ray Bursts an Active Galactic Nuclei studies), ranging from correlations between SNeIa luminosities, progenitor ages, and metallicities, to intrinsic evolution of SNeIa luminosities, to Malmquist bias. However, none of these effects could be convincingly narrowed down as the underlying explanation for the trend. As with the H0LiCOW trend and the trend found by Krishnan \textit{et al.}~\cite{Krishnan:2020obg}, this is one which cannot be fixed by new physics in the early Universe, but only at low redshifts. Finally, while consistent in direction with the previous two trends, the slope in this case is much lower.

Besides these three examples, similar trends have been observed in several other late-time datasets. For instance, Jia, Hu and Wang~\cite{Jia:2022ycc} performed an analysis of the latest Pantheon+ SNeIa sample in combination with BAO and CC data, finding a similar trend of $H_0$ decreasing with increasing redshift, with significance of up to $5.6\sigma$ depending on the binning strategy, and explicitly arguing that this strongly indicates the need for new late-time physics. Importantly, in their analysis Jia, Hu and Wang~\cite{Jia:2022ycc} for the first time removed the correlations between different redshift bins, making use of the transformation matrix discussed by Huterer and Cooray~\cite{Huterer:2004ch}. A similar trend was also found in the Pantheon+ SNeIa sample (without making use of the SH0ES calibrator sample) by Malekjani \textit{et al.}~\cite{Malekjani:2023dky},~\footnote{In this case, rather than binning the data in redshift, the choice was to only use data above a certain redshift and examine the effect of the lower redshift cutoff)} who argued in favor of possible evidence for negative dark energy density (or equivalently $\Omega_m>1$). While not directly focusing on $H_0$ but rather attempting to infer the proper motion of the Solar System from the Pantheon SNeIa, Horstmann, Pietschke and Schwarz~\cite{Horstmann:2021jjg} also found an interesting trend of $H_0$ slightly decreasing with redshift and thereby $\Omega_m$ slightly increasing with redshift (see Fig.~6 in Ref.~\cite{Horstmann:2021jjg}): this inverse correlation is a direct reflection of the $H_0$-$\Omega_m$ degeneracy at the level of low-redshift distances (within $\Lambda$CDM), as one can keep the distance to a certain redshift fixed by decreasing [increasing] one parameter and simultaneously increasing [decreasing] the other. For the same reason, observed increasing trends in $\Omega_m$ from QSOs data (see e.g.\ Refs.~\cite{Risaliti:2018reu,Lusso:2020pdb,Colgain:2022nlb}), which appear to prefer $\Omega \sim 1$ at high redshift, can directly be translated into a decreasing trend in $H_0$ (although not always explicitly reported). A combination of CC, Pantheon SNeIa, and QSOs data was also used by \'{O} Colg\'{a}in \textit{et al.} to report a similar trend~\cite{Colgain:2022rxy}.~\footnote{The reliability of Hubble diagrams constructed out of QSOs data has been questioned (see e.g.\ Refs.~\cite{Velten:2019vwo,Mehrabi:2020zau}), but methods to overcome selection biases and astrophysical evolution in the QSOs parameters have also been tested~\cite{Bargiacchi:2021hdp,Dainotti:2022rfz,Lenart:2022nip,Bargiacchi:2023jse,Dainotti:2023bwq}. Similar considerations hold for Gamma Ray Bursts as a cosmological probe~\cite{Wei:2016gbo,DeSimone:2021rus,Dainotti:2022wli,Dainotti:2022rea,Dainotti:2022ked}.}

Having reported all these trends, the natural questions to ask are to what extent they could have been expected, and what are their implications for the Hubble tension. Let me start from the latter question. Since all these trends were obtained making no assumptions on early-Universe physics, their implication is quite clear: some amount of late-time new physics is clearly needed to ``fix'' these trends, which intriguingly all go in the same direction and show a good compatibility between each other (although not all show the same level of significance). Taking these trends seriously would require this new physics to work in the same direction as that inferred from OAO and CC, i.e.\ in the direction of lowering the expansion rate relative to $\Lambda$CDM at $z>0$, similar to the effect of a phantom component.

As for whether similar trends may be expected even within $\Lambda$CDM, the answer is positive, although somewhat surprising. Na\"{i}vely, one would guess that, because in $\Lambda$CDM DE dies quickly and matter becomes quickly dominant as one moves to high redshift, probes at higher redshift should indicate higher values of $\Omega_m$, and thereby lower values of $H_0$. This expectation, while reasonable, does not appear to be confirmed in analyses of mock data. Specifically, \'{O} Colg\'{a}in \textit{et al.}~\cite{Colgain:2022rxy} and \'{O} Colg\'{a}in, Sheikh-Jabbai and Solomon~\cite{Colgain:2022tql} considered mock distance and expansion rate data matching the expected sensitivity of DESI. After binning this mock data, it was found that as one moves to higher redshifts the $H_0$ and $\Omega_m$ posterior distributions develop non-Gaussian tails in the low $H_0$ and high $\Omega_m$ directions, but the peaks of the distributions actually move in the opposite direction (high $H_0$ and low $\Omega_m$). This was explained in terms of projection effects in combination with the non-Gaussian tails~\cite{Colgain:2022rxy,Colgain:2022tql} (see also Ref.~\cite{Colgain:2023bge}). The surprising conclusion is therefore that one should indeed expect some evolution in $H_0$ and $\Omega_m$, but in the opposite direction compared to what is seen in data. Perhaps more relevant is the fact that the expected size of this ``running'' effect, even at the sensitivity of DESI, is small - far smaller than what is actually observed in data. In fact, as both Fig.~\ref{fig:h0licowdescendingtrend} and Fig.~\ref{fig:baoccsneiadescendingtrend} show, the size of the trend is quite significant even at redshifts $z \lesssim 0.7$, where DE is still dominant and one would not expect any evolution within $\Lambda$CDM, regardless of the direction of the evolution (see e.g.\ Fig.~2 of Ref.~\cite{Colgain:2022rxy}).

Taken in conjunction, all these considerations suggest that the descending trends observed in data are not consistent, neither in terms of size nor direction, with what is expected within $\Lambda$CDM. If we take them seriously, they appear to be the telltale signature of the breakdown of $\Lambda$CDM at late times -- or, to put it differently, a disagreement between mathematics and observations. Importantly, this problem cannot simply be fixed by changing a length scale (such as reducing $r_d$, as is achieved by most early-time new physics models), but require a $w_{\rm eff}(z)$ other than the $\Lambda$CDM one at late times, and therefore call for late-time new physics. Of course, it is also possible that astrophysical systematics may be the cause of these descending trends. One way to eliminate this possibility would be to observe such trends in as many independent datasets as possible, and ascertain whether these independent trends are consistent among each other: should this be the case, it would be extremely hard to make a convincing case for systematics, as different astrophysical systematics would somehow have to conspire to make the trend consistent across independent probes. Current results are indeed moving in this direction, which is why I believe these descending trends should be taken seriously as an indication for new late-time physics.

In closing, I also note that a trend of increasing $S_8$ with redshift was recently reported by Adil \textit{et al.}~\cite{Adil:2023jtu} based on an analysis of $f\sigma_8$ measurements. Moreover, Esposito \textit{et al.}~\cite{Esposito:2022plo} analyzed SZ-identified galaxy cluster number counts from the South Pole Telescope, and Lyman-$\alpha$ spectra from the MIKE/HIRES and X-shooter spectrographs, and inferred a tension between the low value of $S_8$ preferred by low-redshift cluster count data and the high value of $S_8$ inferred from high-redshift Lyman-$\alpha$ data. While the latter was not discussed in light of evolving trends, it can indeed be read in terms of $S_8$ increasing with redshift. If one believes the decreasing $H_0$/increasing $\Omega_m$ trends, this result should not be unexpected, as $S_8 \propto \sqrt{\Omega_m}$, and therefore one should expect the inferred $S_8$ to increase with redshift within $\Lambda$CDM. If substantiated, this trend would suggest that the $H_0$ and $S_8$ tensions are connected, and that further evolution in $S_8$ may be detected with the improvement in precision of WL data.

\subsection{Early integrated Sachs-Wolfe effect}
\label{subsec:eisw}

This fifth hint originates from the following question I first explicitly raised in Ref.~\cite{Vagnozzi:2021gjh}: ``\textit{If solving the Hubble tension requires a significant amount of early-time new physics, why do we not see clear signs of it in CMB data alone}''? Or, equivalently, ``\textit{Why does $\Lambda$CDM appear to fit CMB data so well}''? These queries were driven by the apparently puzzling contradiction between the significant amount of new physics required to solve the Hubble tension, and the lack of detection thereof when analyzing CMB data alone, without the use of any late-time $H_0$ prior. For instance, this is exemplified by the EDE solution to the Hubble tension, which requires $\gtrsim 10\%$ of the energy budget of the Universe at matter-radiation equality to have been in the form of an exotic EDE component, raising the question as to how such a significant amount of new physics may have escaped detection in CMB-only analyses.~\footnote{Tongue-in-cheek, one could say that this is the Hubble tension equivalent of the much more famous Fermi paradox.} As a caveat, hereafter I will focus my discussion on CMB data from \textit{Planck} in order to follow the analysis I performed in Ref.~\cite{Vagnozzi:2021gjh}.

The key observation I made in Ref.~\cite{Vagnozzi:2021gjh} is that for what concerns CMB data, a non-insignificant amount of early-time new physics would first be expected to appear in conjunction with the so-called early integrated Sachs-Wolfe (eISW) effect. The eISW effect is a contribution to CMB anisotropies sourced by time-varying gravitational potentials at early times. In the standard picture, the eISW effect is driven by the fact that the CMB forms when the Universe is not entirely matter dominated, since only in a matter-dominated Universe are gravitational potentials constant in time. The potential decay driven by the non-negligible residual radiation fraction is then responsible for sourcing the eISW effect, which is particularly prominent around the first acoustic peak. To linear order in temperature perturbations, the eISW contribution to the $\ell$th temperature anisotropy multipole, $\Theta_{\ell}(k)$, is given by:
\begin{eqnarray}
\Theta_{\ell}^{\rm eISW}(k) = \int_{0}^{\eta_m}d\eta\,e^{-\tau(\eta)} \left [ \dot{\Psi}(k,\eta)-\dot{\Phi}(k,\eta) \right ]j_{\ell}(k\Delta\eta)\,, \nonumber \\
\label{eq:eisw}
\end{eqnarray}
where $\Psi$ and $\Phi$ are the two scalar potentials in Newtonian gauge, $\eta$ is conformal time  (with $\eta_0$ denoting the conformal time today), $\tau(\eta)$ is the optical depth to a given conformal time, $j_{\ell}$ is the order-$\ell$ spherical Bessel function, $\Delta \eta \equiv \eta-\eta_0$, and $\eta_m$ is the conformal time at an arbitrary point deep inside the matter-domination era (as long as this point is chosen deep inside the matter-domination era, e.g.\ $z \sim 30$, the exact value of $\eta_m$ is irrelevant). The observation I made in Ref.~\cite{Vagnozzi:2021gjh} is then that early-Universe new physics altering the expansion rate around recombination will inevitably modify the evolution of the gravitational potentials, and hence the terms $\dot{\Psi}$ and $\dot{\Phi}$ in Eq.~(\ref{eq:eisw}), which in turn alters the prediction for the eISW contribution to the CMB anisotropies. This would result in a mismatch between the $\Lambda$CDM prediction for the eISW effect, and that of new physics,~\footnote{In particular in Ref.~\cite{Vagnozzi:2021gjh} I argued that models raising the pre-recombination expansion rate, of which EDE can be considered a prototype (but certainly not the only example), naturally boost the eISW effect due to the enhanced potential decay.} which can be tested in a relatively model-agnostic way via an eISW-based consistency test of $\Lambda$CDM (similar in spirit to the OAO test of Sec.~\ref{subsec:oao}, but restricted to early times).

\begin{figure*}
\begin{center}
\includegraphics[height=6.4cm]{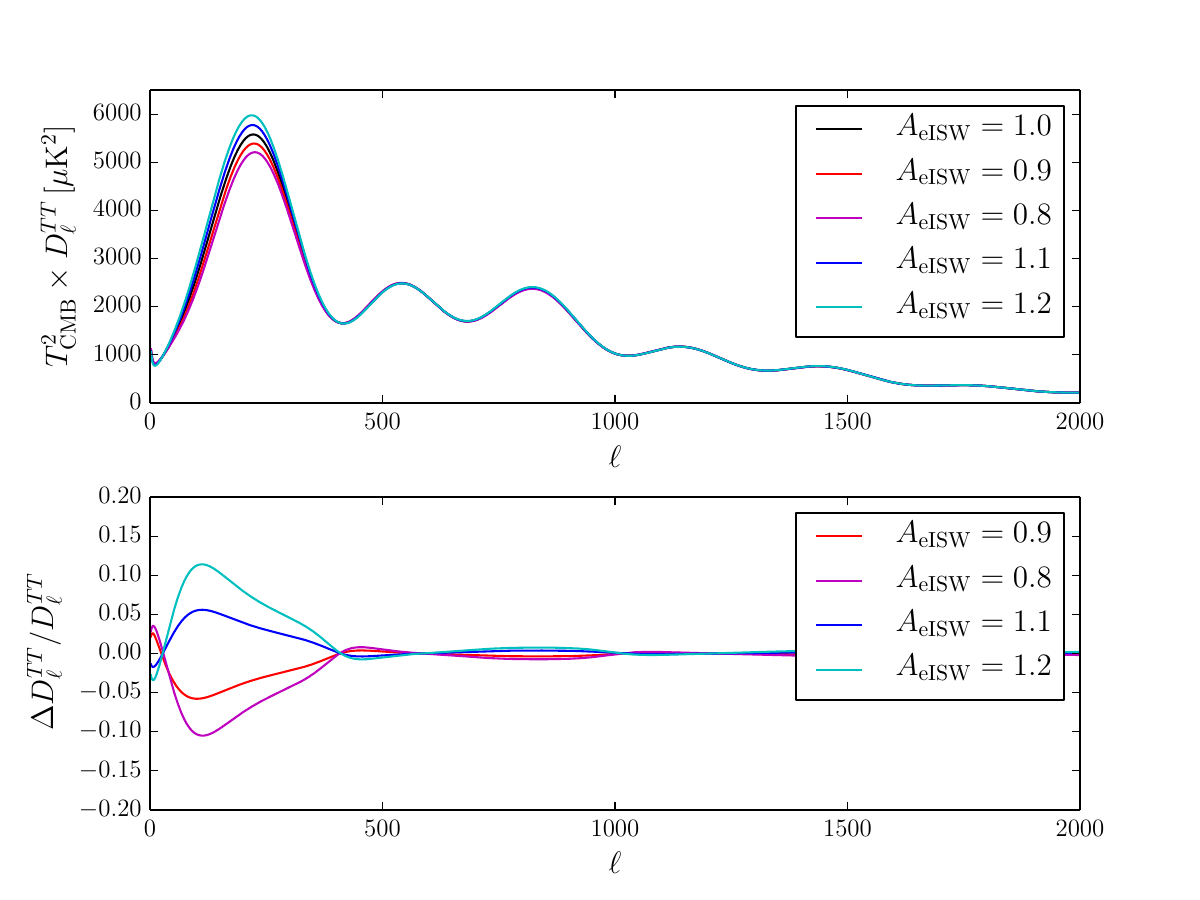}\,\includegraphics[height=6cm]{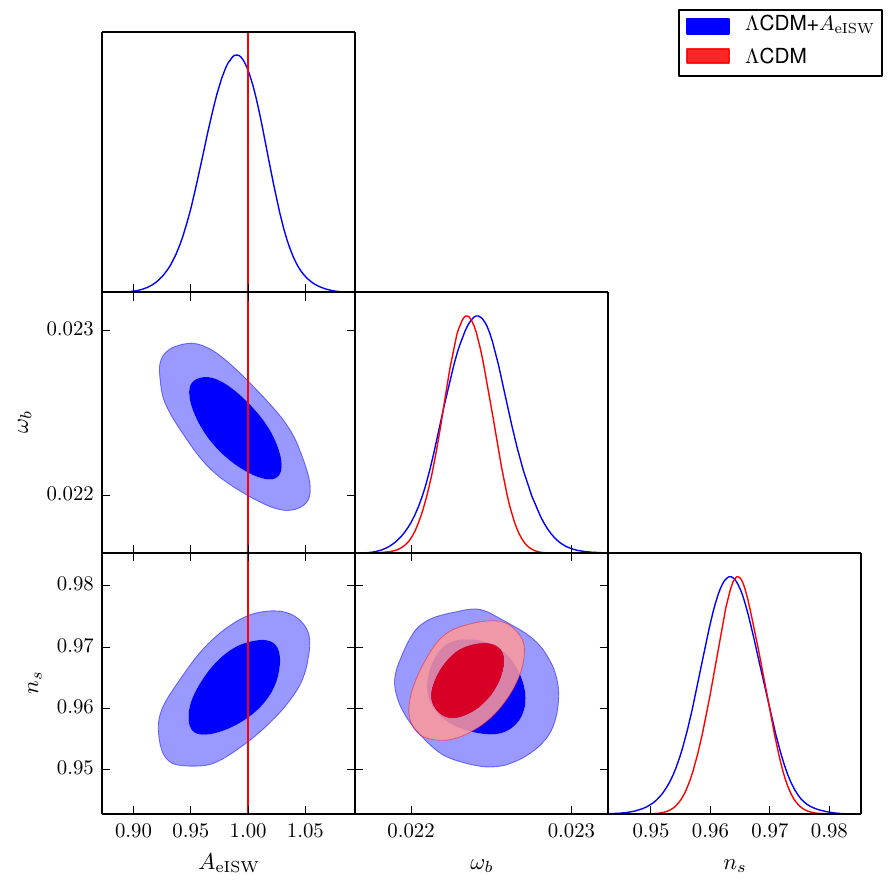}
\caption{\textit{Left panel}: impact of varying $A_{\rm eISW}$ on the CMB temperature power spectrum (upper sub-panel), and relative differences in with respect to the $\Lambda$CDM model (lower sub-panel). \textit{Right panel}: corner plot for $A_{\rm eISW}$, $\omega_b$, and $n_s$, within the $\Lambda$CDM (red contours) and $\Lambda$CDM+$A_{\rm eISW}$ (blue contours) models, from \textit{Planck} data. Reproduced from Fig.~1 and Fig.~2 of Vagnozzi, ``\textit{Consistency tests of $\Lambda$CDM from the early integrated Sachs-Wolfe effect: Implications for early-time new physics and the Hubble tension}'', \textit{Physical Review D}, Volume 104, Issue 6, article number 063524, \href{https://journals.aps.org/prd/abstract/10.1103/PhysRevD.104.063524}{doi:10.1103/PhysRevD.104.063524}, published 15 September 2021~\cite{Vagnozzi:2021gjh}. \textcopyright\,(2021) American Physical Society. Reproduced by permission of the American Physical Society and the author. All rights reserved.}
\label{fig:eisw}
\end{center}
\end{figure*}

To proceed, in Ref.~\cite{Vagnozzi:2021gjh} I isolated the eISW contribution to the CMB power spectra by introducing a phenomenological scaling parameter/fudge factor, the ``eISW amplitude'' $A_{\rm eISW}$, which multiplies Eq.~(\ref{eq:eisw}). This parameter is reminiscent of the phenomenological lensing parameter $A_{\rm lens}$ introduced by Calabrese \textit{et al.}~\cite{Calabrese:2008rt} as a consistency test of the amplitude of CMB lensing (see also Refs.~\cite{Specogna:2023nkq,Giare:2023aix} for recent reassessments pertaining to the $A_{\rm lens}$ problem), and is part of a more general class of phenomenological scaling parameters discussed by Kable, Addison and Bennett~\cite{Kable:2020hcw}, and Ruiz-Granda and Vielva~\cite{Ruiz-Granda:2022bcn}. Note that $A_{\rm eISW}$ should not be viewed as a standard cosmological parameter, but rather as a phenomenological consistency test parameter: $A_{\rm eISW}=1$ corresponds to the prediction of the underlying ``null hypothesis'' model (in this case $\Lambda$CDM), whereas $A_{\rm eISW} \neq 1$ could be an indication for discrepancy at the level of the eISW effect between the underlying model and data. The effect on the CMB temperature power spectrum of varying $A_{\rm eISW}$ is shown in the left panel of Fig.~\ref{fig:eisw}. Clearly, the effect is most prominent around the first peak (with $A_{\rm eISW}>1$/$A_{\rm eISW}<1$ respectively enhancing/reducing the amplitude of the first peak), as one could have expected from purely theoretical considerations~\cite{Vagnozzi:2021gjh}.

In Ref.~\cite{Vagnozzi:2021gjh}, I then fit a 7-parameter $\Lambda$CDM+$A_{\rm eISW}$ cosmological model to \textit{Planck} CMB temperature and polarization data, resulting in the inference of $A_{\rm eISW}=0.988 \pm 0.027$ at 68\%~C.L., in perfect agreement with the $\Lambda$CDM value $A_{\rm eISW}=1$. In addition, the fact that introducing $A_{\rm eISW}$ leads to negligible shifts in the 6 cosmological parameters reinforces the conclusion that $\Lambda$CDM's prediction for the amplitude of the eISW effect agrees perfectly with \textit{Planck} data. The most ``significant''  (to use an euphemism) parameter shifts induced by freeing up $A_{\rm eISW}$ were observed in the scalar spectral index $n_s$ and the physical baryon density $\omega_b$, which as expected are the two parameters most strongly correlated with $A_{\rm eISW}$~\cite{Vagnozzi:2021gjh}: however, both of these parameters shift by less than $\approx 0.3\sigma$ (see the right panel of Fig.~\ref{fig:eisw}). This result was found to be stable against the use of the \texttt{CamSpec} likelihood in place of the \texttt{Plik} one, as well as the inclusion of additional datasets, such as a Big Bang Nucleosynthesis (BBN) prior on $\omega_b$, and constraints on the late-time expansion rate (which can further stabilize constraints on $\Omega_m$) from BAO and uncalibrated SNeIa~\cite{Vagnozzi:2021gjh}. Moreover, the same result was found to be very stable against further parameter space extensions, including cases where the effective number of neutrinos $N_{\rm eff}$, Helium fraction $Y_p$, lensing amplitude $A_{\rm lens}$, running $\alpha_s$ and running of the running $\beta_s$ of the spectral index were allowed to vary~\cite{Vagnozzi:2021gjh}. Therefore, when adopting \textit{Planck} CMB data, the inferred amplitude of the eISW effect is perfectly in agreement with the expectation from $\Lambda$CDM.

I used this result in Ref.~\cite{Vagnozzi:2021gjh} to argue against the presence of a significant amount of early-time new physics which can fit cosmological observations as well as $\Lambda$CDM. Taking as example the EDE solution to the Hubble tension~\cite{Poulin:2018cxd,Kamionkowski:2022pkx,Poulin:2023lkg}, which requires $\gtrsim 10\%$ of the energy budget of the Universe at matter-radiation equality to have been in the form of an exotic EDE component, in Ref.~\cite{Vagnozzi:2021gjh} I explicitly showed that, if all other parameters are kept fixed to their $\Lambda$CDM best-fits, the amplitude of the eISW effect of a Hubble tension-solving EDE cosmology (with $\approx 12\%$ EDE fraction at its maximum) is $\approx 20\%$ higher compared to $\Lambda$CDM, corresponding to $A_{\rm eISW} \simeq 1.20$, and hence completely excluded by the previous analysis. One way of fixing this is to absorb the excess eISW power through parameter shifts, the simplest of which being an increase in the DM density $\omega_c$ (and a further more moderate increase in $n_s$): in fact, such an increase makes matter-radiation equality occur earlier, thereby decreasing the time over which the radiation-driven decay of gravitational potentials can source the eISW effect. However, an increase in $\omega_c$ comes with the deleterious effect of exacerbating the $S_8$ tension, as a larger amount of DM naturally enhances the growth of structure, worsening the tension between CMB and WL probes: this has explicitly been noted in the context of EDE and first emphasized by Hill \textit{et al.}~\cite{Hill:2020osr} (whereas other recent works, such as Goldstein \textit{et al.}~\cite{Goldstein:2023gnw}, discussed related issues pertaining to the scalar spectral index and its inconsistency with measurements from the one-dimensional Lyman-$\alpha$ forest flux power spectrum). I further argued that such a problem, namely an increase in the eISW amplitude which then needs to be compensated in some other way, is not limited to EDE, but quite generally applies to most models which decrease the sound horizon by increasing the pre-recombination expansion rate, i.e.\ a significant fraction of early-time models invoked to solve the Hubble tension~\cite{Vagnozzi:2021gjh}. These models would need to reduce the excess eISW power, the most direct of which involves an increase in $\omega_c$, which has indeed been observed in several early-time models: note that this increase in $\omega_c$ goes precisely in the same direction of the increase in $\Omega_mh^2$ discussed in the context of the second hint (Sec.~\ref{subsec:baodegeneracyslope}), but has a completely different physical origin. At the same time, I caution against generalizing to all early-time models, given that some of these may possess model-dependent ingredients which ameliorate the eISW-related problem (explicitly discussing the cases of Refs.~\cite{Niedermann:2020qbw,Escudero:2021rfi}), and therefore the prospects of specific models should be judged on a case-by-case basis~\cite{Vagnozzi:2021gjh}.

In conclusion the eISW effect, and in particular the high level of consistency between its expected amplitude within the $\Lambda$CDM model and \textit{Planck} CMB data, places important restrictions on the leeway of early-time models which raise the expansion rate around recombination (and thereby enhance the eISW effect), most of which pay the eISW price by increasing $\omega_c$ and thereby worsening the $S_8$ tension. Of course, the conclusions I reached in Ref.~\cite{Vagnozzi:2021gjh} discussed in this hint hinge upon the use of \textit{Planck} CMB data. Although such an analysis has not been explicitly performed in the literature, it is conceivable that the adoption of a different CMB dataset, for instance from the Atacama Cosmology Telescope (ACT), could lead to different conclusions for what concerns $A_{\rm eISW}$. In fact, ACT DR4 data~\cite{ACT:2020gnv} is known to be in slight tension with \textit{Planck} for what concerns the extrapolated height of the first acoustic peak, and thus of parameters whose determination is closely tied to the latter, such as $\omega_b$, $N_{\rm eff}$, and $n_s$~\cite{Handley:2020hdp,Giare:2022rvg,Zhai:2023yny,DiValentino:2023fei,Giare:2023wzl}: by extension, one would expect these conclusions to extend to $A_{\rm eISW}$ as well, although I stress that such a conclusion has yet to be checked explicitly. In fact, if interpreted within an EDE framework, ACT data at face value lead to a detection of non-zero EDE fraction~\cite{Hill:2021yec,Poulin:2021bjr,Smith:2022hwi}, although the ``CMB tension'' between ACT and \textit{Planck} calls for a deeper scrutiny of possible systematics involved. Until these are better understood, or clarified by future CMB data (e.g.\ Refs.~\cite{CMB-S4:2016ple,SimonsObservatory:2018koc,SimonsObservatory:2019qwx}), the most conservative stand one can take is to view \textit{Planck}'s results as placing (upper) guard rails on the amount of early-time new physics which affects the amplitude of the eISW effect.

\subsection{(Fractional) matter density constraints from early-Universe physics insensitive and uncalibrated cosmic standards}
\label{subsec:ucs}

As the reader has hopefully been able to appreciate so far in the context of OAO (Sec.~\ref{subsec:oao}), CC (Sec.~\ref{subsec:cc}), and descending trends in low-redshift data (Sec.~\ref{subsec:trends}), re-analyses of observations relaxing strong assumptions, or alternative analyses which do not depend on these strong assumptions, are very important. If the results of the re-analyses of alternative analyses are consistent with the original/baseline results, new physics attempting to solve tensions by modifying those assumptions cannot be completely successful. In the cases discussed earlier, alternative analyses carrying absolutely no dependence on early-Universe physics systematically uncovered a residual $\approx 2\sigma$ tension between $\Lambda$CDM and late-time probes which cannot, by construction, be fixed by early-time new physics. However, all the resulting constraints were not as competitive as those adopting using CMB data under the assumption of $\Lambda$CDM: this is not surprising, given the enormous constraining power carried by CMB observations, which however have long been thought to depend strongly on early-Universe physics. These observations are the starting point for the discussion of the sixth hint, which relies on a re-analysis of CMB data (in combination with other low-redshift observations) performed by Lin, Chen and Mack~\cite{Lin:2021sfs} with little to no dependence on early-Universe physics, while remaining competitive with the original analysis at least as far as the inferred value of $H_0$ is concerned.

As noted in earlier by Lin, Mack and Hou~\cite{Lin:2019htv}, the Hubble tension is normally presented comparing constraints on $H_0$ only, or sometimes $H_0$ and $r_d$, and this has strenghtened the pre-recombination vs post-recombination narrative. However, besides the age of the Universe $t_U$ as noted elsewhere, a very important role in the Hubble tension discussion can be played by the (fractional) matter density parameter $\Omega_m$, and Lin, Mack and Hou~\cite{Lin:2019htv} advocated for comparing constraints between different probes in the $H_0$-$\Omega_m$ plane, as this allows for a much more complete assessment of the compatibility between different probes, the model-dependence of each, and the ability of non-standard models to truly reconcile all constraints (as referring exclusively to projected information along the $H_0$ direction may obscure the full picture).

CMB observations can set strong constraints on $\Omega_m$, although these have long been thought to depend strongly on the assumed early-Universe physics model (usually $\Lambda$CDM). On the other hand, combined \textit{uncalibrated} BAO and SNeIa data can constrain the $H_0$-normalized background evolution of the late-time Universe: in the case of BAO, this requires treating $r_d$ as a free parameter, which will therefore be completely degenerate with $H_0$.~\footnote{In the case of SNeIa one simply needs to treat the SNeIa absolute magnitude ${\cal M}$ as a free parameter to be marginalized over: however, this is what is routinely done anyway when analyzing uncalibrated SNeIa data (whereas it is much less common to analyze BAO data treating $r_d$ as a free parameter, rather an early-Universe model is usually assumed).} This dataset combination, which essentially constrains $H(z)/H_0$ (and in particular the slope thereof), can constrain $\Omega_m$, although such constraints are not as strong as the model-dependent CMB ones. Finally, there exist plenty of low-redshift probes only involving post-recombination physics (CC are an excellent example), but their precision on $H_0$ is relatively low, because of their own astrophysical uncertainties and their inherent $H_0$-$\Omega_m$ degeneracy. Is it possible to get the best of all three worlds, i.e.\ determinations of $H_0$ and $\Omega_m$ which are as free from early-Universe assumptions as possible, while remaining competitive with the CMB determinations? Lin, Chen and Mack~\cite{Lin:2021sfs} show that this is indeed possible, through a particular combination of the CMB acoustic peak scale $\theta_{\star}$~\footnote{Note that in Ref.~\cite{Lin:2021sfs} and in Fig.~\ref{fig:ucsomegam} $\theta_{\star}$ is denoted by $\theta_{\rm CMB}$.} with uncalibrated BAO and SNeIa data, plus other late-time probes which are completely insensitive to early-Universe physics, which together deliver constraints on $H_0$ and $\Omega_m$ competitive with \textit{Planck}'s $\Lambda$CDM ones. Lin, Chen and Mack~\cite{Lin:2021sfs} denoted this combination ``early Universe physics insensitive and uncalibrated cosmic standards'', or simply ``uncalibrated cosmic standards'' (UCS).

Earlier in Sec.~\ref{subsec:baodegeneracyslope}, I argued that the two distance scales imprinted in CMB and BAO observations, respectively the comoving sound horizon at recombination $r_{\star}$ and at baryon drag $r_d$, are closely related, with $r_d \approx r_{\star}$. This simple observation is the key behind the UCS approach -- Lin, Chen and Mack~\cite{Lin:2021sfs} argued that the difference between the two sound horizons normalized by the Hubble distance, denoted by $\Delta rH_0$, is very small and almost completely insensitive to early-Universe physics:
\begin{eqnarray}
\Delta rH_0 \equiv \left ( r_d-r_{\star} \right ) H_0 = \int_{z_d}^{z_{\star}}dz\,\frac{c_s(z)}{E(z)}\,,
\label{eq:deltar}
\end{eqnarray}
where $c_s(z)$ is the sound speed of the photon-baryon fluid and $E(z) \equiv H(z)/H_0$. The near insensitivity to early-Universe physics of $\Delta rH_0$ is a direct consequence of the narrow separation in redshift $\Delta z_s \equiv z_d-z_{\star} \approx 30$ between the epochs of recombination and baryon drag. On the other hand, the two epochs leave their imprints (respectively in the CMB acoustic peaks and the BAO feature observed in the clustering of the large-scale structure) at widely separated redshifts, $\Delta z \approx 1100$, and this provides the extremely long lever arm required to constrain $\Omega_m$ with a precision comparable to that of CMB data alone when working within a specific model (for instance $\Lambda$CDM).

\begin{figure*}
\begin{center}
\includegraphics[width=0.6\linewidth]{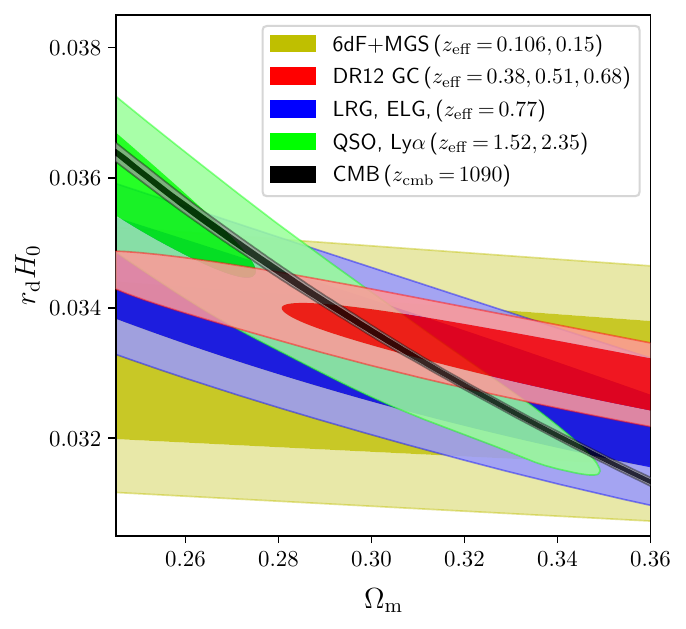}
\caption{Degeneracy contours in the $\Omega_m$-$r_dH_0$ plane from different BAO and CMB (geometrical) probes. Reproduced from Fig.~2 of Lin, Chen \& Mack, ``\textit{Early Universe Physics Insensitive and Uncalibrated Cosmic Standards: Constraints on $\Omega_m$ and Implications for the Hubble Tension}'', \textit{The Astrophysical Journal}, Volume 920, Number 2, article number 159, \href{https://iopscience.iop.org/article/10.3847/1538-4357/ac12cf}{doi:10.3847/1538-4357/ac12cf}, published 25 October 2021~\cite{Lin:2021sfs}. \textcopyright\,(2021) American Astronomical Society. Reproduced by permission of the American Astronomical Society and the authors. All rights reserved.}
\label{fig:degeneracyomegam}
\end{center}
\end{figure*}

\begin{figure*}
\begin{center}
\includegraphics[width=0.7\linewidth]{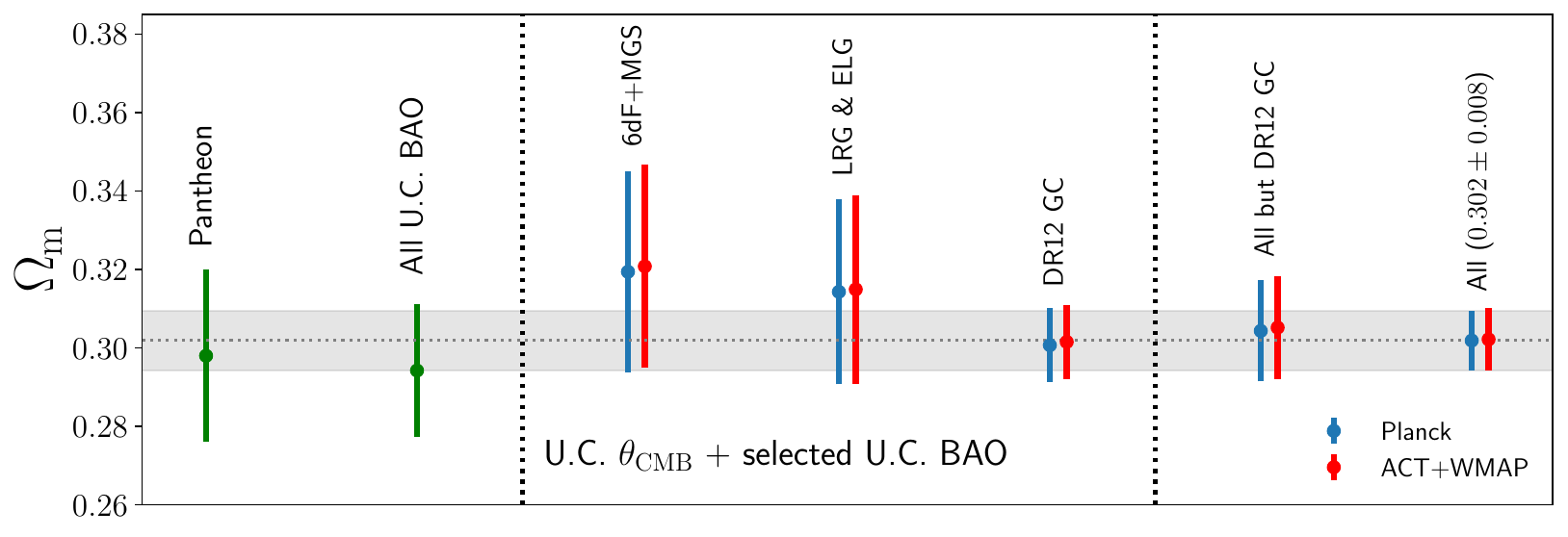}
\caption{Constraints on $\Omega_m$ obtained from different combinations of uncalibrated cosmic standards, with the shaded grey band representing the result to the far right obtained including all UCS measurements. Reproduced from Fig.~3 of Lin, Chen \& Mack, ``\textit{Early Universe Physics Insensitive and Uncalibrated Cosmic Standards: Constraints on $\Omega_m$ and Implications for the Hubble Tension}'', \textit{The Astrophysical Journal}, Volume 920, Number 2, article number 159, \href{https://iopscience.iop.org/article/10.3847/1538-4357/ac12cf}{doi:10.3847/1538-4357/ac12cf}, published 25 October 2021~\cite{Lin:2021sfs}. \textcopyright\,(2021) American Astronomical Society. Reproduced by permission of the American Astronomical Society and the authors. All rights reserved.}
\label{fig:ucsomegam}
\end{center}
\end{figure*}

With these premises, the initial UCS dataset combination considered by Lin, Chen and Mack~\cite{Lin:2021sfs} includes measurements of the CMB acoustic peak scale $\theta_{\star}$ obtained from \textit{Planck} and ACT+WMAP, alongside uncalibrated BAO distance and expansion rate measurements from the 6dFGS, SDSS-MGS, SDSS DR12, and SDSS DR16 galaxy samples, as well as from the eBOSS DR16 quasars and Lyman-$\alpha$ (including their cross-correlations), and finally uncalibrated SNeIa distance moduli measurements from the Pantheon sample. The free parameters varied are $\Omega_m$, the SNeIa absolute magnitude ${\cal M}$, and the combination $r_dH_0$, whereas a BBN prior is set on $\Omega_bh^2$ (required to compute the sound speed of the photon-baryon fluid), and \textit{Planck}-driven priors are set on the redshift of recombination $z_{\star}$ and the redshift separation between recombination and the epoch of baryon drag $\Delta z_s$. The priors on $z_{\star}$ and $\Delta z_s$ fix $\Delta rH_0$ via Eq.~(\ref{eq:deltar}), through which $\theta_{\star}$ can be inferred given the values of $r_dH_0$ and $\Omega_m$: this essentially makes $\theta_{\star}$ another measurement of $\theta_d$ at $z_{\star}$, well above the redshift range probed by BAO measurements ($0 \lesssim z_{\rm obs} \lesssim 2$), providing a powerfully constraining anchor at high redshift and the extremely long lever arm required to tighten constraints on relative changes in $H(z)$ and thereby constrain $\Omega_m$ at the same level as in $\Lambda$CDM, but without assuming any specific early-Universe model. The underlying reason is closely related to our earlier discussion in Sec.~\ref{subsec:baodegeneracyslope} in the context of the $r_d$-$H_0$ degeneracy slopes: in this case, the slope of the $r_dH_0$-$\Omega_m$ degeneracy gets steeper as the redshift of the $\theta_d$ measurements is increased, and it is the fact that $z_{\star} \gg z_{\rm obs}$ which allows for a precise determination of $\Omega_m$, as shown in Fig.~\ref{fig:degeneracyomegam} making use of a number of actual datasets.

From the previous combination and assuming the validity of $\Lambda$CDM post-recombination (but without making any assumption on the pre-recombination expansion history), Lin, Chen and Mack~\cite{Lin:2021sfs} infer $\Omega_m=0.302 \pm 0.008$, only slightly looser than the value obtained from a full analysis of the same data within $\Lambda$CDM (including all CMB information, and not just the location of the first peak), where $\Omega_m=0.310 \pm 0.006$ is inferred. The obtained constraints only makes use of geometrical information on the CMB, and makes almost no assumptions on early-Universe physics, while being also insensitive to degeneracies between $\Omega_m$ and $n_s$ by which a full-shape CMB analysis is instead affected (and which in turn can carry imprints of the slight large-versus-small angular scales discrepancy in \textit{Planck} and therefore potential systematics). Lin, Chen and Mack~\cite{Lin:2021sfs} studied in detail the impact of various assumptions in the choice of UCS datasets and priors (especially for what concerns \textit{Planck} vs ACT+WMAP as well as the inclusion of certain BAO measurements), with the results shown in Fig.~\ref{fig:ucsomegam} indicating that the previous constraint $\Omega_m=0.302 \pm 0.008$ is very robust against these assumptions.

The above analysis delivers a constraint in the $H_0$-$\Omega_m$ plane which is not exactly along the $\Omega_m$ direction, but rather exhibits a small positive correlation described by the following:
\begin{eqnarray}
\left ( \frac{\Omega_m}{0.3} \right ) \left ( \frac{h}{0.7} \right ) ^{-0.08} = 1.0060 \pm 0.0258\,.
\label{eq:principal}
\end{eqnarray}
This suggests that the baseline UCS dataset combination can be further supplemented with additional late-time, but non-local, datasets to break the $H_0$-$\Omega_m$ background evolution degeneracy and thereby determine $H_0$. If these late-time datasets are chosen within the same philosophy of ensuring their independence/insensitivity to early-Universe physics, the resulting $H_0$ inference will be equally insensitive to early-Universe physics and will therefore be extremely valuable in arbitrating the source of the Hubble tension. Lin, Chen and Mack~\cite{Lin:2021sfs} consider many such examples of late-time datasets, ranging from CC, to $\gamma$-ray optical depth measurements, to cosmic age measurements via the ages of old globular clusters (see Sec.~4 of Ref.~\cite{Lin:2021sfs} for further details). While in doing so one is trading the dependence on early-Universe physics for a dependence on other astrophysical effects and thereby potential systematics, the idea is that by choosing a range of late-time measurements as wide and independent as possible one can try to reduce the effects of these systematics as much as possible.

Combining the baseline UCS dataset with all the chosen late-time, non-local observations, Lin, Chen and Mack~\cite{Lin:2021sfs} infer $H_0=(68.8 \pm 1.3)\,{\rm km}/{\rm s}/{\rm Mpc}$, in $2.4\sigma$ tension with the local Cepheid-calibrated SNeIa distance ladder measurement. This inference was obtained by modelling the latest available systematic error estimates for all non-local probes, and was found to be extremely stable against the exclusion of one or more of such probes (most notably the cosmic age estimates, given its dependence on the prior for the formation time of the globular cluster). Remarkably, despite having only assumed the validity of $\Lambda$CDM at late times and introducing almost no early-Universe assumptions, the inferred value of $H_0$ is in excellent agreement with the $\Lambda$CDM-based value obtained from a full analysis of CMB and BAO data, only with uncertainty a factor of $\approx 2$ wider.

What are the implications of these results for the Hubble tension? Clearly the UCS analysis has identified a residual $\gtrsim 2\sigma$ tension which has nothing to do with early-Universe new physics, since virtually any information whatsoever on early-Universe physics has been removed in the analysis. Therefore, such a tension may only be solved by post-recombination non-local new physics (recall that this UCS analysis assumed that the post-recombination expansion history is described by $\Lambda$CDM), and/or local new physics, and/or systematics in either or both the local $H_0$ measurements or the non-local datasets which have been combined with the UCS dataset. However, at least for what concerns the latter, the mutual consistency between the all the UCS+non-local results (see Tab.~1 in Ref.~\cite{Lin:2021sfs} for a complete assessment) suggests that systematics in the latter are unlikely to be an important driver of this residual tension. Leaving aside systematics in general leaves us contemplating either or both late-time (non-local) and local new physics as the most natural avenue for resolving this residual tension. While local new physics would have to go in the direction of lowering the local $H_0$ determinations, late-time (non-local) new physics would once more have to work in the same direction as that inferred from the OAO, CC, and descending trends hints, i.e.\ in the direction of lowering the expansion rate relative to $\Lambda$CDM at $z>0$, similar to the effect of a phantom component. Intriguingly, the size of the ``residual tension'' uncovered by all these probes is consistently in the $2-2.5\sigma$ ballpark which is not high enough to be alarming, but worthy of attention. Overall, the UCS-based analysis strengthens the case for the pre-recombination versus post-recombination narrative not being the end of the story, and strongly makes the case for focusing more on a tension between non-local (pre-recombination \textit{and} post-recombination) and local measurements, in light of the consistency between UCS+non-local measurements and a full $\Lambda$CDM-based CMB analysis.

Finally, one may wonder whether this UCS analysis is truly free of any early-Universe assumption, a question which Lin, Chen and Mack~\cite{Lin:2021sfs} addressed in detail. From the data side, it was argued that the choice of adopting $\Lambda$CDM-based priors for $z_{\star}$ and $\Delta z_s$, as well as the use of the $\Lambda$CDM-based determination of $\theta_{\star}$, have virtually no effect on the results, as does the choice of computing the sound speed $c_s(z)$ as in $\Lambda$CDM once a prior on the physical baryon density is given. From the physics side, Lin, Chen and Mack~\cite{Lin:2021sfs} argued that there are basically three ways the $\theta_d \to \theta_{\star}$ connection, or more precisely the value of $\Delta rH_0$, could come to depend on early-Universe new physics: \textit{a)} some unknown mechanism changing the integrand of $r_{\star}$ relative to $r_d$ (or viceversa) at $z>z_{\star}$ -- else, the fact that the two share the same integrand at this epoch is the reason why the integral in Eq.~(\ref{eq:deltar}) only runs between $z_{\star}$ and $z_d$; \textit{b)} new physics which very significantly affects the normalized expansion rate and/or sound speed in the narrow window $z_d \lesssim z \lesssim z_{\star}$; \textit{c)} a very substantial change in $\Delta z_s=z_{\star}-z_d$. While none of these three ingredients can be definitively excluded, all of them are very hard to achieve, particularly in the context of well-motivated fundamental models. It is more likely that the post-recombination and/or local Universe may need some tweaking to fix the residual early-Universe-independent $2\sigma$ UCS tension.

\subsection{Galaxy power spectrum sound horizon- and equality wavenumber-based determinations of the Hubble constant}
\label{subsec:keq}

As discussed in detail earlier, most attempts at resolving the Hubble tension do so by reducing the sound horizon at recombination $r_{\star}$. The reason is that it is this (acoustic) scale which serves as a standard ruler to calibrate the BAO feature observed both in the CMB and in the clustering of the LSS.~\footnote{In what follows, I will focus on galaxies as tracers of the LSS, and will therefore refer to ``galaxy clustering'', although the subsequent discussion can apply to any LSS tracer.} Indeed, comparing the angular size of the BAO feature extracted from galaxy clustering data to the expected theoretical value of $r_{\star}$ (or more precisely $r_d$, see Sec.~\ref{subsec:baodegeneracyslope} and Sec.~\ref{subsec:ucs}) has been instrumental in ``stabilizing'' CMB-only constraints on $H_0$ and shedding further light on where new physics should act to reduce the Hubble tension. However, $r_s$ is not the only scale imprinted in galaxy surveys, which also contain information on the ``equality scale'' $k_{\rm eq}$, defined as the wavenumber of a perturbation entering the horizon at the epoch of matter-radiation equality at redshift $z_{\rm eq}$. The galaxy power spectrum features a turnaround at $k = k_{\rm eq}$, with modes on smaller scales ($k>k_{\rm eq}$) exhibiting a suppression compared to their large-scale ($k<k_{\rm eq}$) counterparts. The reason is that the $k>k_{\rm eq}$ modes of the DM overdensity field entered the horizon in the radiation domination era, where the large radiation pressure prevented them from growing as much as if they had entered during the matter domination era. This results in a small-scale suppression whose scaling is well approximated as $k^{-4}\ln (k/k_{\rm eq})^2$ (but is highly degenerate with $A_s$ and $n_s$, and therefore relies on a precise determination thereof from CMB data, see Ref.~\cite{Smith:2022iax}), whereas modes on larger scales are unsuppressed and directly trace the primordial spectrum of scalar perturbations presumably set up during inflation.

\begin{figure*}
\begin{center}
\includegraphics[width=0.6\linewidth]{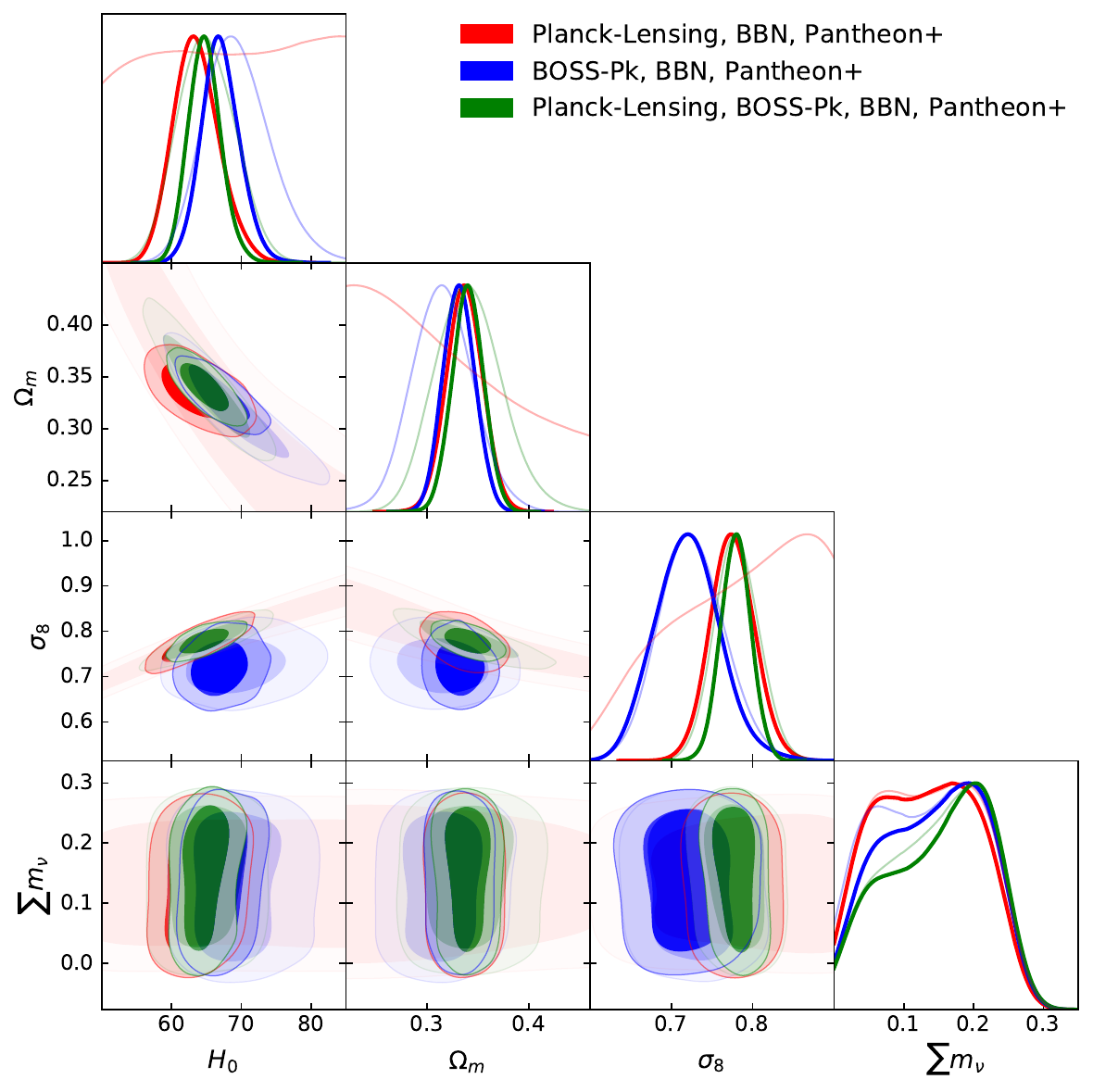}
\caption{Sound horizon-independent constraints on $H_0$ and other parameters from various recent cosmological probes. Reproduced from Fig.~1 of Philcox \textit{et al.}, ``\textit{Determining the Hubble constant without the sound horizon: A  3.6\% constraint on $H_0$ from galaxy surveys, CMB lensing, and supernovae}'', \textit{Physical Review D}, Volume 106, Issue 6, article number 063530, \href{https://journals.aps.org/prd/abstract/10.1103/PhysRevD.106.063530}{doi:10.1103/PhysRevD.106.063530}, published 26 September 2022~\cite{Philcox:2022sgj}. \textcopyright\,(2022) American Physical Society. Reproduced by permission of the American Physical Society and the authors. All rights reserved.}
\label{fig:soundhorizonindependent}
\end{center}
\end{figure*}

The value of the equality scale depends on the balance between energy components with pressure support (typically referred to as ``radiation'') and pressureless matter. This makes the conclusions related to the present hint somewhat dependent on early-time new physics (see also Ref.~\cite{Smith:2022iax}), and this hint should therefore be used more as a consistency test of $\Lambda$CDM rather than a means of excluding specific models of early-time new physics, as I will also discuss in more detail later. With these important caveats in mind, in standard scenarios where no extra components with pressure support are present, the equality scale is given by:
\begin{eqnarray}
k_{\rm eq} = \sqrt{2\Omega_mH_0^2z_{\rm eq}} \,,
\label{eq:keq}
\end{eqnarray}
and can therefore be used as a standard ruler, provided its value can be inferred from galaxy clustering observations. From Eq.~(\ref{eq:keq}) one sees that, since $z_{\rm eq} \propto \Omega_mh^2$, a measurement of $k_{\rm eq}$ in $h\,{\rm Mpc}^{-1}$ is sensitive to the combination $\Gamma \equiv \Omega_mh$. Supplying an external measurement/prior for $\Omega_m$ can therefore break the $\Omega_m-H_0$ degeneracy and lead to an equality-based inference of $H_0$. Besides its intrinsic value as an independent $H_0$ inference, why should this be of particular interest in the context of the Hubble tension? The reason is due to the fact that matter-radiation equality ($z_{\rm eq} \sim 3500$) occurs much earlier than recombination ($z_{\star} \sim 1100$): as a result, most models of early-Universe new physics invoked to solve the Hubble tension predict significant differences at one epoch versus the other (see e.g.\ Fig.~2 of Ref.~\cite{Hill:2020osr} for an example in the context of EDE). This observation provides an avenue for a valuable consistency test of $\Lambda$CDM: since the $r_s$- and $k_{\rm eq}$-based determinations of $H_0$ from galaxy surveys originate from vastly different epochs and underlying physical phenomena, an inconsistency between the two would provide evidence for early-time new physics at redshifts beyond that of recombination, whereas consistency between the two could potentially cause difficulties for models which substantially modify that expansion history immediately prior to recombination (but not necessarily around matter-radiation equality).

After a first exploration in CMB lensing data (where the equality scale enters in a projected way) by Baxter and Sherwin~\cite{Baxter:2020qlr}, the above idea was pursued in detail in two works by Philcox \textit{et al.}~\cite{Philcox:2020xbv,Philcox:2022sgj} in the context of galaxy power spectrum measurements. In particular, Philcox \textit{et al.}~\cite{Philcox:2020xbv,Philcox:2022sgj} developed methods to ``marginalize over the sound horizon'' or, more precisely, ensure that the constraints on $H_0$ resulting from galaxy power spectrum measurements contain as little information about $r_s$ as possible, but result only from $k_{\rm eq}$ information. In the first work, Philcox \textit{et al.}~\cite{Philcox:2020xbv} proceeded by removing any informative prior on $\omega_b$ (e.g.\ from BBN): as the latter is required to calibrate $r_s$, this operation effectively removes any information on $r_s$ at least given the precision of current data, where the $\omega_b$-induced small-scale Jeans suppression cannot be detected at high enough significance to effectively calibrated the sound horizon. Using mock data, Fisher analyses, and scale cuts, the method was demonstrated to efficiently remove sound horizon information in galaxy clustering data, and to capture information from $k_{\rm eq}$ from the full shape of the galaxy power spectrum around the peak (even though the peak itself is difficult to resolve in galaxy surveys). Later Farren \textit{et al.}~\cite{Farren:2021grl} refined this method to more robustly marginalize over $r_s$ and avoid unnecessarily degrading the $H_0$ constraints, by directly marginalizing over templates capturing the power spectrum features related to $r_s$ (i.e.\ the BAO features and Jeans suppression). This allows one to add direct $\omega_b$ information (e.g.\ a BBN prior) while still ensuring that the resulting $H_0$ constraints retain only information from the equality scale.

An application of this method to the latest available cosmological data was presented by Philcox \textit{et al.}~\cite{Philcox:2022sgj}, who considered CMB lensing power spectrum measurements from \textit{Planck}, measurements of the power spectrum monopole, quadrupole, and hexadecapole from the BOSS DR12 galaxy sample, SNeIa measurements from the Pantheon+ sample (in the form of a Gaussian prior on $\Omega_m$), and a BBN prior on $\omega_b$. The constraints from various combinations of these datasets are shown in Fig.~\ref{fig:soundhorizonindependent}. Including the Pantheon+ prior on $\Omega_m$, Philcox \textit{et al.}~\cite{Philcox:2022sgj} find $H_0=64.8^{+2.2}_{-2.5}\,{\rm km}/{\rm s}/{\rm Mpc}$, which shifts (slightly) to $H_0 = 65.0^{+3.9}_{-4.3}\,{\rm km}/{\rm s}/{\rm Mpc}$ when removing this prior. Both inferences are in significant tension with the SH0ES results at the $\gtrsim 3\sigma$ level. Importantly, this residual tension cannot be solved by altering $r_s$ alone, since such a modification would alter the $r_s$-based but not $k_{\rm eq}$-based value of $H_0$. These results were argued to provide an important null test for the $\Lambda$ model at early times (as the $r_s$-free constraints are in excellent agreement with those based on $r_s$ discussed elsewhere), and to disfavor new physics models which affect the sound horizon and equality scale in a very different way. Taken at face value, this suggests the need for new ingredients beyond the ``standard'' $r_s$-reducing ones.

There are, however, a few important caveats, as already alluded to earlier. Firstly, this analysis is mainly a null test of $\Lambda$CDM, and cannot directly be used to exclude specific models. Moreover, the limited sensitivity of current LSS clustering data prevents one from making strong statements in this direction, and implies that priors could play an important role. These points were explored in much more detail by Smith, Poulin and Simon~\cite{Smith:2022iax} where, focusing on three specific beyond-$\Lambda$CDM models, it was shown that:
\begin{itemize}
\item the constraints on $H_0$ obtained from $r_s$-marginalized analyses are model-dependent, although not to a large extent;
\item the results depend to some extent on the assumed priors on $A_s$, $n_s$, and $\Omega_m$;
\item models which introduce additional energy density with significant pressure support (such as EDE or models increasing the effective number of relativistic species $N_{\rm eff}$) can lead to higher $H_0$ values even in $r_s$-free analyses (for such models the $k_{\rm eq}$-based value of $H_0$ is actually slightly lower than the $r_s$-based value, although the error bars are huge).
\end{itemize}
Overall, the conclusion reached by Smith, Poulin and Simon~\cite{Smith:2022iax} is that current $r_s$-free analyses are not strong enough to draw strong conclusions in favor of or against beyond-$\Lambda$CDM models, and therefore that the consistency test proposed by Philcox \textit{et al.}~\cite{Philcox:2022sgj} is inconclusive given the precision of current data, with no strong indication that beyond-$\Lambda$CDM models are in tension with data. However, the forecasts of Philcox \textit{et al.}~\cite{Philcox:2020xbv,Philcox:2022sgj} indicate that the precision of $r_s$-free determinations of $H_0$ will improve significantly with future galaxy clustering data~\cite{LSST:2008ijt,EuclidTheoryWorkingGroup:2012gxx,Spergel:2015sza,DESI:2016fyo,SKA:2018ckk}, and at that point it will be possible to put significant pressure on early-Universe new physics models which alter $r_s$ alone. In fact, it is conceivable that the difference between the $k_{\rm eq}$- and $r_s$-based values of $H_0$, currently consistent within uncertainties, will remain approximately constant, while the error bars will shrink considerably: in this case, the resulting tension between the two determinations may be able to put specific early-time modifications under significant pressure. At the current stage of things, however, drawing strong conclusions does not seem possible, although the consistency (within large error bars) between $k_{\rm eq}$- and $r_s$-based values of $H_0$ certainly disfavors there being an enormous amount of new physics between equality and recombination.

\section{Reflections on promising scenarios moving forward}
\label{sec:promising}

Hopefully by now I have managed to convince the reader that pre-recombination physics \textit{alone} (and I stress that ``alone'' is the keyword here) is not sufficient to solve the Hubble tension: in fact, whatever one does will leave a ``residual'' $1.5$-$2\sigma$ tension which will require new physics at either or both late times and local scales. This has been made especially clear by the first (OAO), third (CC), fourth (descending trends), and sixth (UCS) hints discussed earlier. It is somewhat fair to say that one could have reached such a conclusion empirically: in fact, \textit{in the absence of (high) external local priors on $H_0$}, most early-time modifications struggle to reach $H_0 \sim 71\,{\rm km}/{\rm s}/{\rm Mpc}$, and many actually fall short of the $70$ threshold. Clearly this is far from satisfactory if one is to claim a solution to the Hubble tension. One can of course discuss whether or not it is acceptable or even desirable to include an external $H_0$ prior when assessing the viability of a model in addressing the $H_0$ tension (in which case the previous figures all increase), but I do not believe this is the appropriate place, and I refer the reader to e.g.\ Sec.~VA of Ref.~\cite{Poulin:2023lkg} for a very recent and detailed discussion on the matter. How to make progress from here then? The ``simplest'' scenario which comes to my mind involves a combination of two or three ingredients: early-time new physics, late-time but non-local (i.e.\ cosmological) new physics, and finally possibly local new physics as well. I would expect each of these three ingredients to contribute in a different way. With the caveat that what follows is my opinion and needs to be corroborated by explicit analyses (which I encourage), I now discuss more in detail this scenario.

There is no question that early-time (pre-recombination) new physics would need to do the lion's share of the job in this context, consistently with earlier ``no-go theorems'' excluding late-time new physics alone~\cite{Bernal:2016gxb,Addison:2017fdm,Lemos:2018smw,Aylor:2018drw,Schoneberg:2019wmt,Knox:2019rjx,Arendse:2019hev,Efstathiou:2021ocp,Cai:2021weh,Keeley:2022ojz}. One could envisage considering a particularly successful (relatively speaking) early-time model, such as EDE and variants, varying electron mass (potentially in a curved Universe) resulting in an earlier recombination, self-interacting dark radiation, and so on (see e.g.\ Tab.~1 in Ref.~\cite{Schoneberg:2019wmt}). Empirically, in the absence of external $H_0$ priors, this should bring $H_0$ up to $\sim 70-71\,{\rm km}/{\rm s}/{\rm Mpc}$ in the best-case scenario. In addition to this, a completely decoupled model of late-time (post-recombination, but non-local) new physics could help further push $H_0$ up. In what direction would this late-time new physics need to go? As discussed for several earlier hints, we would require a model which lowers the normalized expansion rate $E(z) \equiv H(z)/H_0$ relative to $\Lambda$CDM. Examples in this sense include phantom DE or phantom-like models: this includes so-called interacting DE models, where non-gravitational interactions between DM and DE result in energy exchange between the two. In the case of energy flow from DM to DE, this results in an effective non-zero positive EoS for DM, while making the DE EoS more negative (i.e.\ moving towards the phantom regime). These models are only mentioned for concreteness, and by no means constitute the only possibilities in this sense.

The question I can already hear the reader asking is whether such a model would even be viable -- after all, isn't it already established that late-time new physics alone cannot solve the Hubble tension~\cite{Bernal:2016gxb,Addison:2017fdm,Lemos:2018smw,Aylor:2018drw,Schoneberg:2019wmt,Knox:2019rjx,Arendse:2019hev,Efstathiou:2021ocp,Cai:2021weh,Keeley:2022ojz}? True, but here I am far from requiring such an ambitious goal for late-time new physics! Rather, all that I require is for the latter to give a further ``push'' in the right direction to the cosmological value of $H_0$, already partially raised by early-time new physics. In other words, I am not asking for late-time new physics to produce a $\Delta H_0 \sim 7\,{\rm km}/{\rm s}/{\rm Mpc}$ (which is definitely not allowed once BAO and Hubble flow SNeIa data are taken into account), but a much more modest value, say $\sim 1.5-2\,{\rm km}/{\rm s}/{\rm Mpc}$ at most.

Something along the lines of what I proposed above should definitely be possible, and let me defend this statement with an empirical argument, in the context of phantom DE. It is often stated that late-time datasets are consistent with a cosmological constant, and do not tolerate large deviations in the DE EoS from $w=-1$. This is broadly speaking true (although it is worth noting that some of the hints for parameter evolution discussed earlier only emerge once one bins data, which is not done in standard analyses), but at the same time \textit{a)} there is certainly room within the error bars for reaching values of $w$ as low as $\sim -1.07$ (or even $\sim -1.10$), and \textit{b)} the central value of the DE EoS $w$ inferred by combining CMB data with several late-time probes which are efficient in breaking the geometrical degeneracy actually falls \textit{slightly} in the phantom regime ($\vert w \vert \sim 1.03-1.05$ or so)~\footnote{See e.g.\ Refs.~\cite{Planck:2018vyg,eBOSS:2020yzd,DAmico:2020kxu,Chudaykin:2020ghx,Brieden:2022lsd,Carrilho:2022mon,Semenaite:2022unt,DES:2018ufa,KiDS:2020ghu,DES:2022ccp,Pan-STARRS1:2017jku,Moresco:2016nqq,Vagnozzi:2020dfn,DiValentino:2020vnx,Yang:2021flj,Moresco:2022phi,Bargiacchi:2021hdp,Grillo:2020yvj,Cao:2021cix,Vagnozzi:2021tjv,Khadka:2020vlh,Khadka:2020tlm,Cao:2021ldv,Khadka:2021xcc,Cao:2023eja} for examples of recent analyses in these directions whose inferred central values of $w$ (in some cases depending on the specific dataset combination or underlying model considered) lie slightly within the phantom regime, arising from a wide range of dataset combinations, mostly involving combinations of \textit{Planck} CMB data with other external late-time measurements, and see Ref.~\cite{Escamilla:2023oce} for a recent reassessment of this point.}-- in this sense, there is already some room within current data for late-time new physics to help in the right direction. Adimensional multipliers quantifying the correlation between $H_0$ and $w$ when confronted against the latest CMB, BAO, and Hubble flow SNeIa data have been estimated in earlier works (see e.g.\ Refs.~\cite{Vagnozzi:2019ezj,deSa:2022hsh}), and indicate that even a modest phantom model with $w \sim -1.04$, by all means tolerated by data, can help as much as $\Delta H_0 \sim 1\,{\rm km}/{\rm s}/{\rm Mpc}$. Still from an empirical point of view, the most successful interacting DE models, once confronted against CMB, BAO, and Hubble flow SNeIa data, have been found to give $\Delta H_0 \sim 1.5\,{\rm km}/{\rm s}/{\rm Mpc}$, which again could be sufficient to bring the Hubble tension to an acceptable level once combined with early-time new physics which has already done the lion's share of the job in raising $H_0$.

The above considerations hinge upon an important point: loosely speaking, it would be desirable for early-time and late-time new physics to contribute ``in phase''/``constructively'' towards $\Delta H_0$. This should be possible if the two models ``decouple'' their tension-solving effects, in other words both raise $H_0$ through completely decoupled physical mechanisms which do not interfere between each other. One might na\"{i}vely guess that early- and late-time new physics should not interfere as they operate at completely different epochs. Such a reasoning however misses one subtle point: parameter shifts induced by one of the two models can limit the tension-solving ability of the other, and viceversa. I already discussed in the second and fifth hints how early-time modifications which only reduce the sound horizon will inevitably induced some parameter shifts. The only way to assess whether these shifts would interfere with the tension-solving ability of additional late-time modifications is to choose a specific combination of models and perform a concrete analysis, and I hope the considerations I have drawn above will encourage others to carry out such an analysis. However, if one were to find two (or more) early- and late-time new physics models which could combine constructively to raise $H_0$, the Hubble tension may well be reduced to an acceptable level, more so than can be achieved with early-time new physics alone (a model which could potentially be interesting in this sense has been presented, albeit in a different context, by Oikonomou~\cite{Oikonomou:2023bah}, based on an axion-Higgs portal and enabling multiple stages of non-standard accelerated expansion both prior to and after recombination, see also Refs.~\cite{Espinosa:2015eda,Dev:2019njv,Im:2019iwd,Bauer:2022rwf,Oikonomou:2023qfz} for related works).

The above considerations, however, may not be the end of the story. In fact, the scenarios proposed so far would have the only effect of raising the \textit{cosmological} value of $H_0$, while leaving the \textit{local} value(s) untouched. In principle there is no reason to entertain such an asymmetry from the model point of view. For instance, what if new physics on local scales were able to reduce the local value of $H_0$? In this case, the cosmological and local values would not need to meet at $\sim 74\,{\rm km}/{\rm s}/{\rm Mpc}$, but perhaps at a slightly lower value, making the task of early-plus-late-time new physics in bringing the Hubble tension to an acceptable level less demanding. It is worth noting that there is a rich literature on local effects which could lead to an overestimate of the distance ladder value of $H_0$, including physical effects such as a local void (the ``Hubble bubble''~\cite{Ding:2019mmw,Cai:2020tpy,Cai:2022dov}), ultra-late transitions in the gravitational constant~\cite{Alestas:2020zol}, screened fifth forces~\cite{Desmond:2019ygn}, and so on. None of these models on their own have proven able to completely address the Hubble tension. In light of the previous discussion, however, such an ambitious goal may no longer be required! Even a small push down by $\vert \Delta H_0 \vert \sim 0.5-1\,{\rm km}/{\rm s}/{\rm Mpc}$ (certainly well within the range of what can be achieved with local new physics!) could be more than sufficient for the combined early-plus-late-plus-local new physics scenario to bring the Hubble tension to an acceptable level. In passing let me mention that a breakdown of the Friedmann-Lema\^{i}tre-Robertson-Walker framework, and more generally of the cosmological principle, may be a scenario particularly worthy of consideration.

\begin{figure*}
\begin{center}
\includegraphics[width=0.7\linewidth]{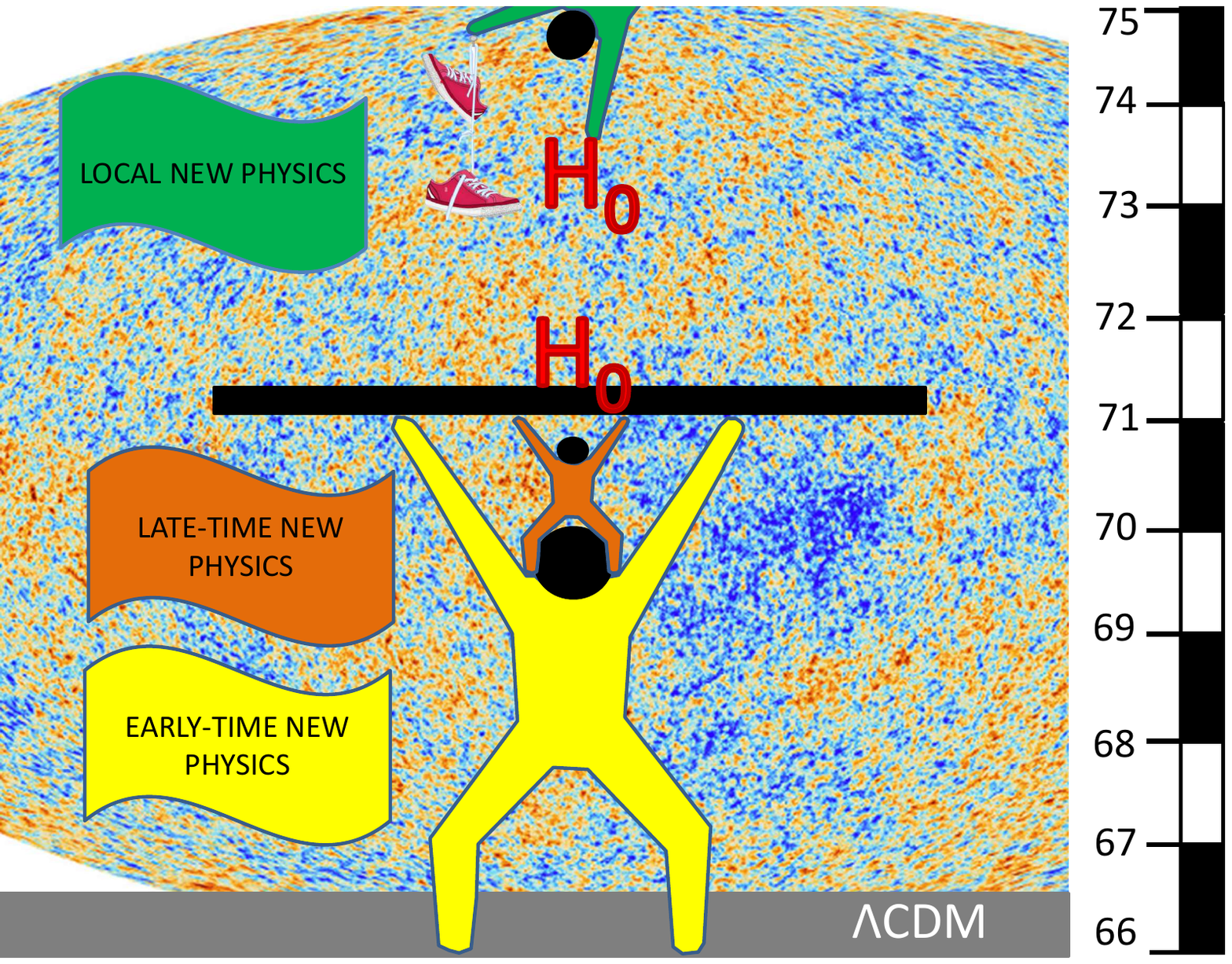}
\caption{Pictorial representation of my proposed approach towards promising solutions to the Hubble tension, discussed in Sec.~\ref{sec:promising}. The grey band at the bottom of the picture represents the $\Lambda$CDM baseline. The yellow person illustrates the effect of early-time new physics, which is still required to do the lion's share of the job. A smaller but non-negligible contribution from late-time new physics, represented by the orange person, pushes the cosmological value of $H_0$ further up. Finally, local new physics, corresponding to the green person holding the SH0ES, can help bringing the local value of $H_0$ slightly down, and make the job of new cosmological physics less demanding. Picture drawn by Cristina Ghirardini.}
\label{fig:pictorial}
\end{center}
\end{figure*}

The combined early-plus-late-plus-local new physics scenario I have discussed in words above is perhaps best summarized pictorially in Fig.~\ref{fig:pictorial}. I hope that this helps convey the idea that early-time new physics (the yellow person) still needs to carry the lion's share of the task, but with a little help from late-time new physics (the orange person) and local new physics (the green person with the SH0ES), the task of solving the Hubble tension or at least bringing it to an acceptable level may be made less daunting. There are three obvious objections to this otherwise arguably nice picture:
\begin{itemize}
\item \textit{What about Occam's razor}? True, such a scenario may be viewed as unnecessary complicated, and aesthetically unpleasing, to the eyes of some. My view on this point is that, in the field of cosmology, Occam's razor (and with it the concept of Bayesian evidence) is sometimes overused/abused. Not always is the simplest model the ``most correct'' one just because it fits the data better or with less parameters. Think, for example, about the many parameters for which we need to impose physical priors in cosmological analyses, to avoid inferring unphysical values thereof (the sum of the neutrino masses is an excellent example), which would otherwise be preferred by the data. If the true model chosen by Nature is actually as complicated as shown in Fig.~\ref{fig:pictorial}...then so be it! With a bit of poetic liberty, allow me to paraphrase Neil de Grasse Tyson and state that ``Nature [the universe] is under no obligation to appear beautiful or simple in our eyes [make sense to you]''.
\item \textit{You haven't mentioned any concrete combination of early-plus-late-plus-local new physics models which does what you advocate}. True once more, and this is left as an exercise to the reader and to aspiring Hubble tension solvers (including myself -- so I hope to report on this in future work).
\item \textit{Can all these early-plus-late-plus-local new physics ingredients come from one single underlying microphysical model, thus reducing the overall complexity}? Maybe, why not? But once more, I am leaving this as an exercise to the reader and to aspiring Hubble tension solvers (this time likely not including myself).
\end{itemize}
Granted, the above objections remain, but I hope the reader now finds them less worrisome.

Before concluding, let me address the remaining elephant in the room: the $S_8$ tension. It would be preferable if the scenario depicted in Fig.~\ref{fig:pictorial} were to ameliorate the $S_8$ tension as well. It is now becoming increasingly clearer to the community that simultaneously addressing the $H_0$ and $S_8$ tensions requires multiple modifications to $\Lambda$CDM at multiple epochs, not unlike the picture shown in Fig.~\ref{fig:pictorial}. In particular, it is likely that addressing the $S_8$ tension will require some physical mechanism to suppress small-scale power~\cite{Amon:2022azi}, rather than playing around with $\Omega_m$ (although complete consensus has yet to be reached on this matter). It would of course be desirable if the physical mechanism responsible for ameliorating the $S_8$ tension does not do so at the expense of the $H_0$ tension, i.e.\ does not interfere with the tension-solving abilities of the other models. In this respect, there is a class of models I find extremely interesting, which I will call ``dark scattering'' models.

The dark scattering class of models involves elastic (pure momentum) scattering-type interactions between dark components (DM, DE, or some other additional dark component), or between dark components and baryons.~\footnote{Dark scattering-type scenarios studied in the literature include DM-DE scattering~\cite{Simpson:2010vh,Xu:2011nr,Richarte:2014yva,Boehmer:2015sha,Tamanini:2015iia,Koivisto:2015qua,Pourtsidou:2016ico,Dutta:2017kch,Kumar:2017bpv,Linton:2017ged,Bose:2017jjx,Bose:2018zpk,Asghari:2019qld,Kase:2019veo,Kase:2019mox,Chamings:2019kcl,Asghari:2020ffe,Amendola:2020ldb,DeFelice:2020icf,BeltranJimenez:2020qdu,Figueruelo:2021elm,BeltranJimenez:2021wbq,Carrilho:2021hly,Linton:2021cgd,Mancini:2021lec,Carrilho:2021rqo,Poulin:2022sgp,Cardona:2022mdq,Piga:2022mge,Jimenez:2022evh}, DM-photon scattering~\cite{Wilkinson:2013kia,Stadler:2018jin,Kumar:2018yhh,Yadav:2019jio}, DM-neutrino scattering~\cite{Serra:2009uu,Wilkinson:2014ksa,Escudero:2015yka,DiValentino:2017oaw,Stadler:2019dii,Hooper:2021rjc}, DM-baryon scattering~\cite{Dvorkin:2013cea,Gluscevic:2017ywp,Boddy:2018kfv,Xu:2018efh,Boddy:2018wzy,Ali-Haimoud:2021lka,Buen-Abad:2021mvc,Nguyen:2021cnb,Rogers:2021byl,Driskell:2022pax,He:2023dbn}, DM self-scattering and scattering with dark radiation~\cite{Foot:2014uba,Foot:2014osa,Cyr-Racine:2015ihg,Vogelsberger:2015gpr,Foot:2016wvj,Archidiacono:2017slj,Buen-Abad:2017gxg,Archidiacono:2019wdp}, ``multi-interacting DM'' scenarios featuring multiple similar interactions simultaneously~\cite{Becker:2020hzj}, and DE-baryon scattering~\cite{Vagnozzi:2019kvw,Jimenez:2020ysu,Vagnozzi:2021quy,Benisty:2021cmq,Ferlito:2022mok}.} The attractive feature of these interactions is that, at linear level in cosmological perturbations, they only affect the evolution of perturbations, but not that of the background. In other words, these interactions only exchange momentum, but not energy (in an appropriate reference frame, e.g.\ the frame where one of the two scattering fluids is at rest). Such a feature is attractive because it can ensure that whatever mechanism is responsible for suppressing small-scale power does not alter the background expansion, and therefore the quality of the fit to BAO and Hubble flow SNeIa data. While this is not a guarantee of such a model not interfering with the tension-solving abilities of whatever other ingredients are solving the Hubble tension, it can certainly be a fairly important step in the right direction. Therefore, even if the scenario depicted in Fig.~\ref{fig:pictorial} were not to solve the $S_8$ tension, one could envisage adding further scattering-type ingredients to ameliorate the situation, ideally without significantly affecting the successes obtained with $H_0$. For instance, Poulin \textit{et al.}~\cite{Poulin:2022sgp} recently proposed DM-DE scattering as a potential solution to the $S_8$ tension, precisely due to the features discussed above. It would be interesting to further combine this model with a decoupled successful early-time model, e.g.\ EDE, to assess the potential of this combination in solving both the Hubble and $S_8$ tensions without the two models ``interfering distructively''.

Is it natural, however, to consider ingredients which only modify the evolution of perturbations but not the background? There is, in fact, a very natural example of such an ingredient within the $\Lambda$CDM model: Thomson scattering between photons and baryons. At the level of first-order Boltzmann equations, Thomson scattering only affects the evolution of perturbations, and more precisely the photon and baryon velocity divergences $\theta_{\gamma}$ and $\theta_b$ (without affecting the photon and baryon overdensities $\delta_{\gamma}$ and $\delta_b$), whereas the background expansion remains unchanged while the two components scatter, with their energy densities evolving as $\rho_{\gamma} \propto (1+z)^4$ and $\rho_b \propto (1+z)^3$ respectively. In the case of dark components, it has actually been argued that at linear order in perturbations a scenario with only momentum exchange can actually be constructed even at the microphysical level by a particular choice of interaction Lagrangian, which within the classification of coupled DE models first discussed by Pourtsidou, Skordis and Copeland~\cite{Pourtsidou:2013nha} and later by Skordis, Pourtsidou and Copeland~\cite{Skordis:2015yra}, was termed \textit{Type 3 model}: this model involves couplings of the covariant derivative of the DE scalar field $\phi$ to the velocity field of the scattering species and was shown to lead to a coupling current vector whose time component, in the DE rest frame, is null, ensuring the absence of energy exchange (thus featuring pure momentum exchange) at least up to linear order. This is of course no guarantee that the resulting model is well-motivated from the microphysical point of view, but is certainly of interest from the perspective of model-builders.

\section{Conclusions}
\label{sec:conclusions}

I have argued the Hubble tension, which at the time of writing remains unsolved, will ultimately require more than just early-time new physics alone. My claim is not based on a proof in a strict sense, but rather on a number of independent hints, at first glance somewhat unrelated, but which paint a more coherent picture once viewed from a holistic perspective. In a nutshell, these seven hints can be summarized as follows:
\begin{enumerate}
\item the $z \lesssim 10$ $\Lambda$CDM Universe appears a bit too young to accommodate the oldest astrophysical objects at high redshift, and this is a problem which cannot be fixed by new physics in the early Universe, but requires new physics at late times or in the local Universe;
\item early-Universe new physics which only reduces the sound horizon cannot simultaneously agree with CMB, BAO, local $H_0$, and WL data, and will necessarily introduce new tensions (or worsen existing ones) involving some of these observations;
\item cosmic chronometers show a residual $\approx 2\sigma$ tension with local $H_0$ measurements within $\Lambda$CDM, a conclusion which is completely independent of early-Universe physics, and which therefore cannot be addressed by invoking the latter;
\item if the $H_0$ tension is physical (in the sense of not being due to systematics) \textit{and} calls for some amount of late-time new physics, evolving $H_0$ trends should be seen at intermediate redshifts between the CMB and local scales, and by now several independent hints thereof have appeared;
\item the early ISW effect places very restrictive guard rails on what early-Universe physics may or may not do, and for models enhancing the pre-recombination expansion rate this often results in fixing the otherwise overpredicted eISW amplitude at the expense of worsening other tensions;
\item early-Universe-independent uncalibrated cosmic standard constraints on $\Omega_m$ and $H_0$ show a residual $\approx 2\sigma$ tension with local $H_0$ measurements which cannot, by construction, be fixed by early-time new physics, but most involve late-time or local new physics;
\item the good agreement between sound horizon- and equality wavenumber-based constraints on $H_0$ from galaxy power spectra measurements makes it relatively unlikely that a significant amount of new physics operating before recombination can be at play.
\end{enumerate}

As I stressed earlier \textit{alone} is the keyword here, as there is no question (given the important role of BAO and Hubble flow SNeIa data) that early-time new physics will have to play an important role in solving the Hubble tension. However, on its own, I have argued that early-time new physics will not be able to achieve the task. I believe that an ultimately promising scenario may be the one depicted in Fig.~\ref{fig:pictorial}: early-time new physics doing an important job in raising $H_0$ can further be helped (in a smaller proportion) by late-time new physics, and by local new physics which lowers the local value of $H_0$, easing the task of ensuring the cosmological and local values of $H_0$ meet halfway. It would be intriguing if the late-time contribution were also to help with the $S_8$ tension without affecting the Hubble tension-solving ingredients of the early-time part. In this sense, I have argued that late-time dark scattering-type models with pure momentum exchange, which to linear order in perturbations do not alter the background evolution, are particularly promising and worth exploring much more in the context of cosmological tensions.

Despite significant efforts from the theoretical and observational sides, the solution to the Hubble tension has so far eluded us. However, I believe there is reason to be optimistic, especially in light of upcoming Stage-IV cosmological data which may contain key evidence for beyond-$\Lambda$CDM physics. These may potentially confirm some or all of the hints I have presented here which, I hope, can play a small role in guiding the community towards a building a new concordance model of cosmology.

\begin{acknowledgments}
\noindent I thank Cristina Ghirardini for producing Fig.~\ref{fig:pictorial}. I acknowledge support from the Istituto Nazionale di Fisica Nucleare (INFN, National Institute for Nuclear Physics) through the Iniziativa Specifica ``FieLds And Gravity'' (FLAG). This publication is based upon work from the COST Action CA21136 ``Addressing observational tensions in cosmology with systematics and fundamental physics (CosmoVerse), supported by COST (European Cooperation in Science and Technology).
\end{acknowledgments}

\bibliography{sevenhints.bib}

\end{document}